\documentclass[twocolumn,journal, bookmarks=false]{IEEEtran}
\IEEEoverridecommandlockouts

\usepackage{color}
\usepackage{graphicx}
\usepackage{amsmath}
\usepackage{amssymb}
\usepackage{algorithm}
\usepackage{algorithmic}
\usepackage{amsmath}
\usepackage{multirow}
\usepackage{booktabs}
\usepackage{array}
\usepackage{amsthm}
\usepackage{stfloats}
\usepackage{caption}
\usepackage{subfigure}
\usepackage{bm}
\usepackage{epstopdf}
\usepackage{setspace}
\usepackage{gensymb}
\usepackage{url}
\usepackage{verbatim}
\usepackage{balance}

\usepackage{hyperref}
\usepackage{cleveref} 

\hypersetup{colorlinks = true, 
    linkcolor = black, 
    urlcolor = black,
    citecolor = black} 

\newcommand{\be}{\begin{equation}}
        \newcommand{\ee}{\end{equation}}
\newcommand{\bea}{\begin{eqnarray}}
        \newcommand{\eea}{\end{eqnarray}}
\newcommand{\ba}{\begin{array}}
        \newcommand{\ea}{\end{array}}

\allowdisplaybreaks[4]

\title{MIMO-OFDM ISAC Waveform Design for Range-Doppler Sidelobe Suppression
    \thanks{A portion of this paper was presented at the  IEEE International Conference on Communications, 2024 \cite{LPS2024}.}
    \thanks{P. Li and M. Li are with the School of Information and Communication Engineering, Dalian University of Technology, Dalian 116024, China (e-mail: lipeishi@mail.dlut.edu.cn; mli@dlut.edu.cn).}
    \thanks{R. Liu and A. Lee Swindlehurst are with the Center for Pervasive Communications and Computing, University of California, Irvine, CA 92697, USA (e-mail: rangl2@uci.edu; swindle@uci.edu).}
    \thanks{Q. Liu is with the School of Computer Science and Technology, Dalian University of Technology, Dalian 116024, China (e-mail: qianliu@dlut.edu.cn).}
}
\author{Peishi Li,~\IEEEmembership{Graduate Student Member,~IEEE,}
    Ming Li,~\IEEEmembership{Senior Member,~IEEE,}
    Rang Liu,~\IEEEmembership{Member,~IEEE,}  
    \\ 
    Qian Liu,~\IEEEmembership{Member,~IEEE,}
    and A. Lee Swindlehurst, ~\IEEEmembership{Fellow,~IEEE}}

\begin{document}

\maketitle
\thispagestyle{empty}
\begin{abstract}
Integrated sensing and communication (ISAC) is a key enabling technique for future wireless networks owing to its efficient hardware and spectrum utilization. In this paper, we focus on dual-functional waveform design for a multi-input multi-output (MIMO) orthogonal frequency division multiplexing (OFDM) ISAC system, which is considered to be a promising solution for practical deployment. Since the dual-functional waveform carries communication information, its random nature leads to high range-Doppler sidelobes in the ambiguity function, which in turn degrades radar sensing performance. To suppress range-Doppler sidelobes, we propose a novel symbol-level precoding (SLP) based waveform design for MIMO-OFDM ISAC systems by fully exploiting the temporal degrees of freedom (DoFs). Our goal is to minimize the range-Doppler integrated sidelobe level (ISL) while satisfying the constraints of target illumination power, multi-user communication quality of service (QoS), and constant-modulus transmission. To solve the resulting non-convex waveform design problem, we develop an efficient algorithm using the majorization-minimization (MM) and alternative direction method of multipliers (ADMM) methods. Simulation results show that the proposed waveform has significantly reduced range-Doppler sidelobes compared with signals designed only for communications and other baselines. In addition, the proposed waveform design achieves target detection and estimation performance close to that achievable by waveforms designed only for radar, which demonstrates the superiority of the proposed SLP-based ISAC approach.
\end{abstract}

\vspace{1mm}
\begin{IEEEkeywords}
    Integrated sensing and communication (ISAC), waveform design, range-Doppler sidelobes, symbol-level precoding (SLP), MIMO-OFDM systems.
\end{IEEEkeywords}

\vspace{1mm}
\section{Introduction}

Driven by the rapid evolution of and burgeoning demand for advanced intelligent applications such as autonomous driving, extended reality (XR), and smart cities, integrated sensing and communication (ISAC) has gained widespread attention as a crucial enabler for future sixth-generation (6G) wireless networks \cite{background_1}-\cite{background_3}. Unlike traditional designs that consider sensing and communication as two separate functions, ISAC technology ingeniously integrates them within a unified system architecture, leveraging resource sharing across temporal, frequency, and spatial domains to improve spectrum efficiency,  reduce hardware cost, and achieve cooperation gain \cite{ISAC_power}, \cite{ISAC_power2}.

For wireless communication functions, orthogonal frequency division multiplexing (OFDM) has been widely adopted in various wireless systems, such as 5G NR \cite{5GNR} and IEEE 802.11 \cite{wifi}, owing to its advantages of high spectrum efficiency, robustness to frequency selective fading, and implementation flexibility. Furthermore, for radar sensing functions, the multi-carrier structure of OFDM signals can ensure the independence and orthogonality of spectral components, which enables range and velocity estimation to be performed in independent dimensions \cite{no_coupling}, \cite{OFDMradar}.
The feasibility and effectiveness of using OFDM signals for target detection and parameter estimation have already been substantiated \cite{sturm}.
Therefore, OFDM signals are considered to be highly suitable as ISAC dual-functional waveforms in light of their potential for excellent sensing and communication performance as well as their considerable flexibility.
Moreover, the use of multi-input multi-output (MIMO) antenna arrays provides additional spatial degrees of freedom (DoFs) that can be exploited to achieve beamforming gains, spatial multiplexing, and waveform diversity for both communication and radar sensing systems \cite{MIMO_RC}. Consequently, employing MIMO-OFDM hardware architectures and signal waveforms to realize both communication and sensing functions is one of the most promising approaches for ISAC deployment in practical wireless networks.

The crucial challenge for MIMO-OFDM ISAC lies in the dual-functional waveform and beamforming designs that must balance the wireless communications and radar sensing performance requirements \cite{ISAC_waveform_design}. To achieve a satisfactory performance trade-off, numerous ISAC beamforming designs have been proposed to exploit the spatial DoFs based on different communication and sensing performance metrics. Typical sensing performance metrics include the signal-to-interference-plus-noise ratio (SINR) at the radar receiver \cite{sinr}, the mean squared error (MSE) between the actual and desired beampattern \cite{mse}, the Cram\'{e}r-Rao bound (CRB) \cite{CRB} for the target parameter estimates, etc. On the other hand, commonly used communication performance metrics are the achievable rate \cite{rate}, the SINR of the communication users \cite{commSINR1}, and the level of multi-user interference (MUI) \cite{MUI}.

The beamforming designs cited above solely focus on the spatial second-order statistics of the transmit signals, and do not take into consideration the temporal characteristics of the waveform or the influence of the random information symbols on radar sensing performance. In particular, the performance of a radar system heavily depends on the {\em ambiguity function}, or time-frequency autocorrelation, of the transmit waveform. To obtain an ideal thumbtack-like ambiguity function, various deterministic sequences with favorable temporal correlation properties, such as the Zadoff-Chu (ZC) sequence \cite{chu}, are widely adopted in radar system design \cite{detection_AF}. On the other hand, communication signals are designed to be as random as possible to increase information transfer, and when used in an ISAC system, such signals result in high sidelobes in the ambiguity function over the range-Doppler plane \cite{range_sidelobe_randomness}, \cite{range_sidelobe_randomness2}. Since the ambiguity function represents the output of the matched filter utilized for echo signal processing, high sidelobes lead to the masking of weak targets and increased false alarms for ``ghost'' targets \cite{sidelobe_detect}.

There are two typical approaches for addressing the influence of random information symbols on the radar sidelobes. First, instead of using matched filtering, \cite{sturm} introduced a reciprocal filtering-based method that performs element-wise division in both the subcarrier and symbol domains at the OFDM radar receiver. If the radar receiver has full knowledge of the transmitted symbols, this step can effectively remove the impact of the random symbols. However, reciprocal filtering can enhance the noise power, which will in turn deteriorate the radar sensing performance \cite{spectral_division}. As a second approach, \cite{radar_beam} proposed to transmit a dual-functional waveform composed of both precoded random communication symbols and deterministic radar probing sequences. The two signals with different functions and temporal characteristics are combined together via beamforming by exploiting the distinct spatial channels of the targets and users. The beamforming mitigates interference between the communication and sensing signals and can balance the performance trade-off. Despite the simplicity of the concept, the overall effectiveness of this approach is not as effective as that of an integrated design. In particular, the interference between the communication and radar signals cannot be entirely eliminated in complex scenarios where, for example, the target and user are near each other. Under these circumstances, the radar echo will still be influenced by the random communication symbols, which will lead to high sidelobe levels and deteriorate the radar sensing performance. 
In summary, state-of-the-art ISAC waveform designs do not effectively address the impact of communication signal randomness on the range-Doppler sensing performance, which is a crucial factor for radar detection and range/velocity estimation.

In this paper, we investigate a promising alternative technique based on the recently emerged concept of symbol-level precoding (SLP) to tackle the conflicts between radar sensing and communication in ISAC systems. SLP is a non-linear precoding approach that exploits knowledge of not only the spatial channels of the users, but also the symbols transmitted to them. The precoding is calculated in each symbol time slot and provides additional DoFs for waveform design in both temporal and spatial domains \cite{SLP_DoF}. In particular, for the ISAC application, SLP enables the base station (BS) to more flexibly manipulate the temporal characteristics of the dual-functional waveform, and consequently suppress the sidelobes of the ambiguity function over the range-Doppler plane, thereby significantly mitigating the masking effect and improving the radar sensing performance. 
In addition, from the perspective of multi-user communications, SLP technique can exploit the symbol and channel state information to convert the harmful multiuser interference (MUI) into constructive interference (CI), thus achieving a better communication quality of service (QoS) \cite{SLP_DoF2}. Thus, SLP is an ideal candidate for ISAC dual-functional waveform design owing to its potential to improve the performance of both communication and sensing functions by exploiting additional DoFs available in the spatial and temporal domains. 

Inspired by the potential of SLP in terms of its temporal and spatial design flexibility, this paper presents the first investigation of SLP-based ISAC waveform design for range-Doppler sidelobe suppression in the MIMO-OFDM ISAC systems, where a multi-antenna dual-function BS simultaneously performs both downlink multi-user communication and radar sensing functions. The main contributions of the paper are summarized as follows:
\begin{itemize}
    \item To investigate the impact of the transmit waveform on radar sensing performance, we formulate the discrete periodic ambiguity function for the MIMO-OFDM signal, which provides a quantitative basis for evaluating the radar sensing performance in MIMO-OFDM ISAC systems and an appropriate metric for waveform designs.

    \item To achieve the goal of range-Doppler sidelobe suppression, we introduce a novel SLP-based MIMO-OFDM ISAC waveform design with the aim of minimizing the range-Doppler ISL of the ambiguity function, while satisfying the constraints on the target illumination power, the multi-user communication QoS, and the constant-modulus of the transmitted waveform. To solve the resulting non-convex waveform design problem, an efficient algorithm is developed with the aid of the majorization-minimization (MM) and alternative direction method of multipliers (ADMM) methods. 
        
    \item We provided extensive simulation results to demonstrate the superiority of the proposed SLP-based ISAC waveform design and highlight the resulting radar sensing performance improvement. The proposed waveform can remarkably reduce the normalized range-Doppler ISL by more than $45$dB compared with waveforms designed only for the communication users. Owing to the sidelobe suppression, the proposed waveform design can achieve excellent target detection and estimation performance close to that of waveforms designed only for radar functionality, thereby significantly enhancing the sensing capability for weak targets.
\end{itemize}

\textit{Notation:} Unless otherwise specified, the following notation is used throughout the paper. Boldface lower-case letters (e.g., $\mathbf{x}$) indicate column vectors, while bold upper-case letters (e.g., $\mathbf{X}$) indicate matrices. The sets $\mathbb{C}$ and $\mathbb{Z}$ represent the collection of complex numbers and integers, respectively. Superscripts $()^{\ast}$, $()^{T}$, $()^{H}$, and $()^{-1}$ indicate the conjugate, transpose, transpose-conjugate, and inverse operations, respectively. The operators $\mathfrak{R}\{ \cdot \}$ and $\mathfrak{I}\{ \cdot \}$ extract the real and imaginary parts of a complex number. An $N\times N$ identify matrix is denoted by $\mathbf{I}_{N}$, while $\mathbf{1}_{N}$ and $\mathbf{0}_{N}$ are $N \times 1$ vectors with all-one or all-zero entries, respectively. The element-wise absolute value and $\ell_{2}$ norm of a vector are respectively indicated by $| \cdot |$ and $\| \cdot \|$. The function $\text{Tr}\{ \mathbf{X} \}$ is the trace of the square matrix $\mathbf{X}$, $\text{vec} \{ \mathbf{X} \}$ vectorizes the matrix $\mathbf{X}$ column-by-column, and $\text{mat}_{N \times M}\{ \mathbf{y} \}$ denotes the $N \times M$ matrix satisfying $\text{vec}\{ \text{mat}_{N \times M}\{ \mathbf{y} \} \} = \mathbf{y}$. The operators $\otimes$ and $\odot$ represent the Kronecker product and Hadamard (element-wise) product, respectively, and $\angle a$ is the angle of complex-valued $a$. The notation $\mathbf{A}(i, :)$, $\mathbf{A}(:, j)$, and $\mathbf{A}(i, j)$ indicates the $i$-th row, the $j$-th column, and the $(i, j)$-th entry of matrix $\mathbf{A}$, respectively. The integer part of a number is represented by 
$\lfloor\cdot\rfloor$, and $\text{mod}(m,n)$ returns the remainder after division of $m$ by $n$.

\vspace{2mm}
\section{System Model and Problem Formulation} \label{sec:system model}
\begin{figure}[!t]
    \centering
    \includegraphics[width = 3 in]{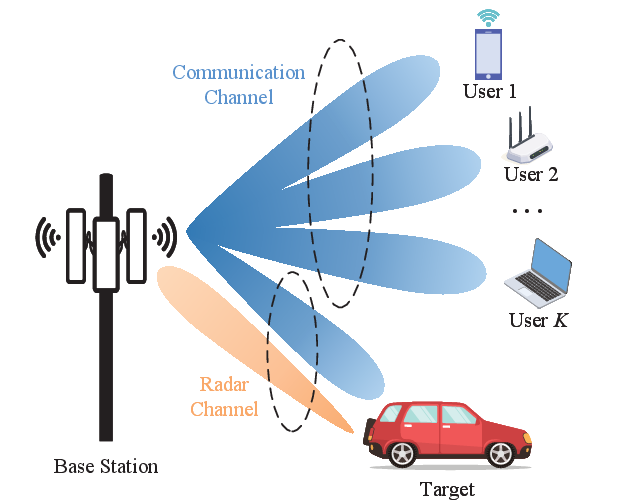}
    \caption{Illustration of an MIMO-OFDM ISAC system.}
    \label{fig:system_model}
\end{figure}

The considered MIMO-OFDM ISAC system is depicted in Fig. \ref{fig:system_model}, where a BS employs $N_{\text{t}}$ transmit antennas and $N_{\text{r}}$ receive antennas arranged as uniform linear arrays (ULAs). The dual-function BS emits an OFDM signal with $N_{\text{c}}$ subcarriers and $N_{\text{s}}$ symbols to serve $K$ single-antenna communication users and simultaneously performs radar sensing using the echo signals to detect prospective targets and estimate their range and velocity. We denote $\mathcal{N}_{\text{t}} = \{ 1, 2, \dots, N_{\text{t}} \}$, $\mathcal{N}_{\text{r}} = \{ 1, 2, \dots, N_{\text{r}} \}$, $\mathcal{K} = \{ 1, 2, \dots, K \}$, $\mathcal{N}_{\text{c}} = \{ 0, 1, \dots, N_{\text{c}}-1 \}$, and $\mathcal{N}_{\text{s}} = \{ 0, 1, \dots, N_{\text{s}}-1 \}$ as the index sets of the transmit antennas, receive antennas, users, subcarriers, and symbol time-slots, respectively. We assume that the target of interest is located at azimuth $\theta_0$, range $R_0$, and moves with velocity $v_0$. In the following, we will introduce the comprehensive signal model and radar echo signal processing for the MIMO-OFDM ISAC system, as well as the performance metrics we use for evaluating the radar sensing and communication.

\subsection{Transmit Signal Model}
The modulated symbol vector on the $n$-th subcarrier during the $m$-th time-slot is denoted as $\mathbf{s}_{n, m} \triangleq \left[ s_{n, m,1},  \dots, s_{n,m,K} \right]^T \in \mathbb{C}^{K}, n \in \mathcal{N}_{\text{c}}, m \in \mathcal{N_{\text{s}}}$, where each symbol is drawn from a fixed constellation set. Without loss of generality, we assume that all users employ an identical modulation scheme. To constructively exploit the multi-user interference as well as suppress the range-Doppler sidelobes of the ambiguity function, the dual-function BS employs a non-linear SLP technique that takes into consideration the modulated symbols within each time-slot and their temporal properties across time-slots.

Let $\mathbf{x}_{n, m} \in \mathbb{C}^{N_{\text{t}}}$ denote the precoded signal on the $n$-th subcarrier during the $m$-th time-slot, which carries information symbols $\mathbf{s}_{n, m}$. 
With the precoded signal sequences $\mathbf{x}_{n, m}$, $n \in \mathcal{N}_{\text{c}}$, $m \in \mathcal{N_{\text{s}}}$, for $N_{\text{s}}$ symbol time-slots, we can perform OFDM modulation and obtain the baseband time domain OFDM signal $\widetilde{\mathbf{x}}(t) \in \mathbb{C}^{N_{\text{t}}}$, which can be mathematically expressed as
\begin{equation} \label{eq:baseband_OFDM}
    \begin{aligned}
        \widetilde{\mathbf{x}}(t) = \frac{1}{\sqrt{N_{\text{c}}}}  \sum_{m = 0}^{N_{\text{s}}-1} \sum_{n = 0}^{N_{\text{c}}-1}  \mathbf{x}_{n, m} e^{\jmath 2 \pi n \Delta f t_m} \text{rect} \bigg( \frac{t_m}{T_{\text{tot}}} \bigg),
    \end{aligned}
\end{equation}
where we define the following variables: $\Delta f = 1/T$ is the subcarrier spacing, $T$ is the OFDM symbol duration, $T_{\text{CP}}$ is the cyclic prefix (CP) duration, $T_{\text{tot}} = T + T_{\text{CP}}$ is the total symbol duration including the CP, and $t_m=t- m T_{\text{tot}}$ is the relative fast-time variable during the $m$-th OFDM symbol. The function $\text{rect}(\cdot)$ is a rectangular pulse defined as
\begin{equation}
    \text{rect} \Big( \frac{t}{T} \Big) \triangleq \begin{cases}
        1 & ~~0 \leq t \leq T; \\
        0 & ~~\text{otherwise}.
    \end{cases}
\end{equation}
Before transmission, the baseband OFDM signal in (\ref{eq:baseband_OFDM}) is up-converted to the radio frequency (RF) domain, and the up-converted OFDM signal can be written as
\begin{equation}
    \widetilde{\mathbf{x}}_{\text{RF}}(t) = \widetilde{\mathbf{x}}(t) e^{\jmath 2 \pi f_{\text{c}} t},
\end{equation}
where $f_{\text{c}}$ is the carrier frequency.

\subsection{Radar Echo Signal Model and Signal Processing}
The transmitted OFDM signal $\widetilde{\mathbf{x}}_{\text{RF}}(t)$ is reflected by the target, and the echo signal is then received by the BS. 
For simplicity, we assume that the radar cross section (RCS) of the target is constant during the total OFDM signal duration $N_{\text{s}} T_{\text{tot}}$ \cite{constant_RCS}. Moreover, let $\tau_0 = 2 R_0/c_0$ denote the round-trip delay and $f_{\text{d}} = 2 v_0 f_{\text{c}}/c_0$ the Doppler shift, where $c_0$ is the speed of light. 
Then, after down-conversion, the baseband echo signal received by the $N_{\text{r}}$ antennas at the BS can be written as \cite{radar_echo1}, \cite{radar_echo2}
\begin{subequations}
    \begin{align}
        \!\widetilde{\mathbf{y}}(t)
         & = \alpha_0 \mathbf{a}_{\text{R}}(\theta_0) \mathbf{a}_{\text{T}}^H \! (\theta_0)\widetilde{\mathbf{x}}(t \!- \! \tau_0) e^{-\jmath2 \pi f_{\text{c}} \tau_0} e^{\jmath 2\pi f_{\text{d}}t} \!+\! \widetilde{\mathbf{n}}(t)    \label{eq:radar_signal_a} \\
         & =  \frac{\bar{\alpha}_0}{\sqrt{N_{\text{c}}}} \mathbf{a}_{\text{R}}(\theta_0) \mathbf{a}_{\text{T}}^H (\theta_0) \sum_{m = 0}^{N_{\text{s}}-1} \sum_{n = 0}^{N_{\text{c}}-1} \mathbf{x}_{n, m} e^{\jmath 2 \pi n \Delta f (t_m-\tau_0)} \notag     \\
         & \hspace{2cm} \times e^{\jmath 2\pi f_{\text{d}} t } \text{rect} \bigg( \frac{t_m -\tau_0 }{T_{\text{tot}}} \bigg) + \widetilde{\mathbf{n}}(t), \label{eq:radar_signal_b}
    \end{align}
\end{subequations}
where $\alpha_0 \triangleq \sqrt{\frac{\sigma_{\text{RCS}} \lambda^2}{(4\pi)^3 R_0^4}}$ denotes the attenuation factor including the path loss and the target RCS, $\sigma_{\text{RCS}}$ is the RCS of the target, $\lambda = c_0 / f_{\text{c}}$ is the wavelength, and the vector $\widetilde{\mathbf{n}}(t) \in \mathbb{C}^{N_{\text{r}}}$ represents additive white Gaussian noise (AWGN) with zero mean and variance $\sigma_{\text{r}}^2$. Since the first exponential term in (\ref{eq:radar_signal_a})  is a constant phase shift, we denote $\bar{\alpha}_0=\alpha e^{-j2 \pi f_{\text{c}} \tau_0}$ for brevity. The vectors $\mathbf{a}_{\text{T}}(\theta_0)$ and $\mathbf{a}_{\text{R}}(\theta_0)$ represent the transmit and receive steering vectors, and are given by
\begin{subequations}
    \begin{align}
        \mathbf{a}_{\text{T}}(\theta_0) & \triangleq \big[ 1, e^{\jmath 2 \pi  \sin(\theta_0) \frac{d_{\text{T}}}{\lambda} }, \dots, e^{\jmath 2 \pi (N_{\text{t}}-1) \sin(\theta_0) \frac{d_{\text{T}}}{\lambda} }  \big]^T \\
        \mathbf{a}_{\text{R}}(\theta_0) & \triangleq \big[ 1, e^{\jmath 2 \pi  \sin(\theta_0) \frac{d_{\text{R}}}{\lambda}}, \dots, e^{\jmath 2 \pi (N_{\text{r}}-1) \sin(\theta_0) \frac{d_{\text{R}}}{\lambda} }  \big]^T,
    \end{align}
\end{subequations}
where $d_{\text{T}}$ and $d_{\text{R}}$ are respectively the transmit and receive antenna spacing.

After sampling the radar echo signal in (\ref{eq:radar_signal_b}) at the sampling rate $f_{\text{s}}=N_{\text{c}} \Delta f$, the resulting discrete-time radar echo signal can be written as
\begin{equation} \label{eq:discrete_y}
    \begin{aligned}
        \widetilde{\mathbf{y}}_{p, m}
         & = \bar{\alpha}_0 \mathbf{a}_{\text{R}}(\theta_0) \mathbf{a}_{\text{T}}^H(\theta_0) \sum_{n=0}^{N_{\text{c}}-1}  \mathbf{x}_{n, m} e^{\jmath 2 \pi n \frac{p}{N_{\text{c}}}} e^{-\jmath 2 \pi n \Delta f \tau_0} \\
         & ~~~~ \times  e^{\jmath 2\pi \frac{p}{N_{\text{c}}}  f_{\text{d}} T  } e^{\jmath 2\pi m f_{\text{d}} T_{\text{tot}} } + \widetilde{\mathbf{n}}_{p,m} , ~~ p \in \mathcal{N}_{\text{c}},
    \end{aligned}
\end{equation}
where $\widetilde{\mathbf{y}}_{p, m}$ denotes the $p$-th sample of the $m$-th OFDM symbol in $\widetilde{\mathbf{y}}(t)$, and $\widetilde{\mathbf{n}}_{p,m} = \widetilde{\mathbf{n}}(p \frac{T}{N_{\text{c}}} + m T_{\text{tot}})$. To provide more insight, we analyze the effect of each exponential term in~(\ref{eq:discrete_y}) below:
\begin{itemize}
    \item The first exponential term $e^{\jmath 2 \pi n p/N_{\text{c}}}$ represents the effect of OFDM modulation.
    \item The second exponential term $e^{-\jmath 2 \pi n \Delta f \tau_0}$ can be interpreted as the delay-induced phase shift over the OFDM subcarriers, which is used for range estimation. Note that the randomness of the precoded signal $\mathbf{x}_{n, m}$ across the subcarriers leads to high range sidelobe levels.
    \item The third and fourth exponential terms, $e^{\jmath 2\pi p  f_{\text{d}} T /N_{\text{c}} }$ and  $e^{\jmath 2\pi m f_{\text{d}} T_{\text{tot}} }$, denote the Doppler-induced phase shift in the fast- and slow-time domains, respectively. The former can cause inter-carrier interference (ICI), while the latter can be utilized for velocity estimation. Analogously, the randomness of the precoded signal $\mathbf{x}_{n, m}$ in the slow-time domain results in high Doppler sidelobes.
\end{itemize}

To avoid ICI, the OFDM subcarrier spacing $\Delta f$ is typically chosen to be larger than the Doppler shift $f_{\text{d}}$, i.e. $f_{\text{d}} T= f_{\text{d}} / \Delta f \ll 1$. Thus, the Doppler-induced phase shift $e^{\jmath 2 \pi p  f_{\text{d}} T/ N_{\text{c}}}$ within one symbol duration can be omitted using a suitable parametrization \cite{ICI_omit}. To estimate the range and velocity of the target, we perform the discrete Fourier transform (DFT) over each OFDM symbol and convert the discrete-time signal $\widetilde{\mathbf{y}}_{p,m}$ into the frequency domain, which results in the following expression:
\begin{subequations}
    \begin{align}
        \!\!\!\mathbf{y}_{n, m}
         & = \bar{\alpha}_0 \mathbf{a}_{\text{R}}\!(\theta_0) \mathbf{a}_{\text{T}}^H \! (\theta_0) \mathbf{x}_{n, m} e^{\jmath 2 \pi (f_{\text{d}} m T_{\text{tot}} - n \Delta f \tau_0)} \!+\! \mathbf{n}_{n,m} \\
         & = \bar{\alpha}_0 \mathbf{a}_{\text{R}}(\theta_0) \mathbf{a}_{\text{T}}^H (\theta_0) \mathbf{x}_{n, m} e^{\jmath 2 \pi (\nu_0 \frac{m}{N_{\text{s}}}-  l_0\frac{n}{N_{\text{c}}})} +\mathbf{n}_{n,m},
    \end{align}
\end{subequations}
where $\mathbf{y}_{n, m} \in \mathbb{C}^{N_{\text{r}}}$ represents the discrete echo signal on the $n$-th subcarrier during the $m$-th OFDM symbol. We assume that the round-trip delay $\tau_0$ and the Doppler shift $f_{\text{d}}$ are integer multiples of the delay and Doppler resolutions \cite{delay_inter}, i.e.,
\begin{equation}
    \tau_0 = \frac{l_0}{N_{\text{c}} \Delta f}, ~~~~ f_{\text{d}} = \frac{\nu_0}{N_{\text{s}} T_{\text{tot}}},
\end{equation}
where $l_0$ and $\nu_0$ represent the unknown range and Doppler bins, respectively. The vector $\mathbf{n}_{n,m}$ represents the Fourier transform of the noise $\widetilde{\mathbf{n}}_{p,m}$.

We assume that the target azimuth angle $\theta_0$ is known at the BS and focus on estimating the target range and velocity. Thus, after spatial filtering at the BS using the steering vector $ \mathbf{a}_{\text{R}}(\theta_0) $, the output signal can be expressed as
\begin{subequations} \label{eq:y0}
    \begin{align}
        y_{n, m}
         & = \frac{1}{N_{\text{r}}} \mathbf{a}_{\text{R}}^H(\theta_0) \mathbf{y}_{n,m}                                                                                 \\
         & = \bar{\alpha}_0 \mathbf{a}_{\text{T}}^H (\theta_0) \mathbf{x}_{n, m} e^{-\jmath 2 \pi l_0\frac{n}{N_{\text{c}}}} e^{\jmath 2\pi \nu_0 \frac{m}{N_{\text{s}}}} \notag \\
         & \hspace{2.3cm} + \frac{1}{N_{\text{r}}} \mathbf{a}_{\text{R}}^H(\theta_0) \mathbf{n}_{n,m}.
    \end{align}
\end{subequations}
Then, the range- and Doppler-induced phase rotations $e^{\!-\jmath 2 \pi l_0\frac{n}{N_{\text{c}}}}\!$ and $e^{\jmath 2\pi \nu_0 \frac{m}{N_{\text{s}}}}$ can be used to estimate the range and velocity of the target, respectively. To describe the signal processing steps more concisely, (\ref{eq:y0}) is reformulated in matrix notation as
\begin{equation}
    \mathbf{Y} = \bar{\alpha}_0 \mathbf{D}_{l_0} \overline{\mathbf{X}} \mathbf{D}_{\nu_0} +  \mathbf{N},
\end{equation}
where we define the following matrices:
\begin{subequations}
    \begin{align}
         & \mathbf{Y}(n, m)  \triangleq y_{n, m}, \\
         & \mathbf{X}(1+n N_{\text{t}} : (n+1) N_{\text{t}}, m) \triangleq \mathbf{x}_{n,m},                                                                                                \\
         & \overline{\mathbf{X}} \triangleq (\mathbf{I}_{N_{\text{c}}} \otimes \mathbf{a}^H_{\text{T}}(\theta_0) ) \mathbf{X},\\     
         & \mathbf{D}_{l_0} \triangleq \text{diag}\big( 1, e^{-\jmath2 \pi l_0 \frac{1}{N_{\text{c}}}}, \dots, e^{-\jmath2 \pi l_0 \frac{N_{\text{c}}-1}{N_{\text{c}}}} \big),  \label{eq:D_l_define} \\
         & \mathbf{D}_{\nu_0} \triangleq \text{diag}(1, e^{\jmath2 \pi \nu_0 \frac{1}{N_{\text{s}}}}, \dots, e^{\jmath 2 \pi \nu_0  \frac{N_{\text{s}}-1}{N_{\text{s}}}}),  \label{eq:D_nu_define}\\
          & \mathbf{N}(n, m)  \triangleq \mathbf{a}_{\text{R}}^H(\theta_0) \mathbf{n}_{n,m} / N_{\text{r}}.
    \end{align}
\end{subequations}

Classical OFDM radar signal processing performs the Hadamard product of $\mathbf{Y}$ and $\overline{\mathbf{X}}^{\ast}$, then applies the inverse-DFT (IDFT) and DFT over the OFDM subcarriers and symbols to obtain the range-Doppler map \cite{PMF-CC}, \cite{MF_PMF-CC}. This matched filtering operation can be mathematically expressed as
\begin{equation} \label{eq:RDM}
    \boldsymbol{\chi}=  \mathbf{F}^{H}_{N_{\text{c}}} (\mathbf{Y} \odot \overline{\mathbf{X}}^{\ast}) \mathbf{F}_{N_{\text{s}}},
\end{equation}
where $\boldsymbol{\chi} \in \mathbb{C}^{N_\text{c} \times N_\text{s}}$ is the range-Doppler map, and $\mathbf{F}_{N_\text{c}} \in \mathbb{C}^{N_\text{c} \times N_\text{c}}$ denotes the normalized DFT matrix. Target detection and range-velocity parameter estimation can then be performed by searching for peaks in the range-Doppler map, e.g., using a constant false alarm rate (CFAR) detector. Consequently, high-quality radar sensing is greatly dependent on achieving low range-Doppler sidelobe levels as well as low noise power levels (i.e. high SNR). In the following subsection, we will introduce two metrics for quantifying the range-Doppler sidelobe levels and the receive SNR, upon which the waveform design can be based.

\vspace{1mm}
\subsection{Radar Performance Metric}
We see from (\ref{eq:RDM}) that the sidelobes of the range-Doppler map are essentially determined by the transmit waveform $\mathbf{X}$, and the effectiveness of the matched filter output can be evaluated by examining the ambiguity function of the transmit waveform $\mathbf{X}$~\cite{AF_PMFCC}.
The range-Doppler sidelobes of the ambiguity function is a crucial metric for evaluating target detection and parameter estimation performance in radar signal analysis and waveform designs \cite{AF_sidelobe}. For example, a low range-Doppler sidelobe level will make the radar less prone to false alarms or detection errors~\cite{range_sidelobe}. Thus, we will employ the ambiguity function of the transmit waveform to analyze the characteristics of the range-Doppler sidelobes and design the waveform to suppress them. 
In the following, we will derive expressions for the ambiguity function of the MIMO-OFDM signal and the corresponding range-Doppler ISL, which will be our metric of choice for quantifying the sidelobe performance.

The ambiguity function is essentially the time-frequency composite auto-correlation function of the transmitted signal, which is defined as \cite{AF_radar}
\begin{equation} \label{eq:AF_define}
    \chi(\tau,f_{\text{d}}) \triangleq \int_{-\infty}^{\infty} \widetilde{x}_0(t) \widetilde{x}_0^{\ast}(t+\tau)e^{\jmath 2 \pi f_{\text{d}}t} \text{d}t,
\end{equation}
where $\widetilde{x}_0(t)=\mathbf{a}_{\text{T}}^H(\theta_0) \widetilde{\mathbf{x}}(t)$ is the OFDM signal after beamforming to the known azimuth angle $\theta_0$. In this paper, we assume that the round-trip delay is smaller than the CP duration. Consequently, we can justifiably use the discrete periodic ambiguity function of the OFDM signal to simplify the derivations \cite{discreteAF}, \cite{PAF}, which is expressed as
\begin{equation} \label{eq:PAF_define}
    \!\! \chi (l, \nu) \!=\! \!\sum_{m = 0}^{N_{\text{s}}-1} \!\sum_{p = 0}^{N_{\text{c}}-1}\! \mathbf{a}_{\text{T}}^H \!(\theta_0) \widetilde{\mathbf{x}}_{p,m} \big(\mathbf{a}_{\text{T}}^H\!(\theta_0) \widetilde{\mathbf{x}}_{p+l,m} \big)^{\ast} e^{\jmath 2 \pi \nu \frac{m}{N_{\text{s}}}},
\end{equation}
where $l \in \mathcal{N_{\text{c}}}$ and $\nu \in \mathcal{N_{\text{s}}}$ are the indices of the range and Doppler bins, respectively. The vector $\widetilde{\mathbf{x}}_{p,m}$ represents the $p$-th sample of the transmitted time-domain OFDM signal $\widetilde{\mathbf{x}}(t)$, and is given by
\begin{equation} \label{eq:x_0pm}
    \widetilde{\mathbf{x}}_{p,m} = \frac{1}{\sqrt{N_{\text{c}}}} \sum_{n=0}^{N_{\text{c}}-1} \mathbf{x}_{n,m} e^{\jmath 2 \pi p\frac{n}{N_{\text{c}}}}.
\end{equation}
Substituting (\ref{eq:x_0pm}) into (\ref{eq:PAF_define}), the discrete periodic ambiguity function of the OFDM signal can be formulated as
\begin{subequations}
    \begin{align}
        \!\!\chi(l, \nu)
         & =  \frac{1}{N_{\text{c}}} \sum_{m = 0}^{N_{\text{s}}-1} \sum_{p = 0}^{N_{\text{c}}-1} \sum_{n = 0}^{N_{\text{c}}-1} \big(\mathbf{a}_{\text{T}}^H (\theta_0) \mathbf{x}_{n, m} \big) \big(\mathbf{a}_{\text{T}}^H (\theta_0) \mathbf{x}_{n, m} \big)^{\ast}         \notag  \\
         & \hspace{3cm} \times e^{-\jmath 2 \pi l\frac{n}{N_{\text{c}}}} e^{\jmath 2\pi \nu \frac{m}{N_{\text{s}}}}                                                                                                                                                       \\
         & = \sum_{m = 0}^{N_{\text{s}}-1} \sum_{n = 0}^{N_{\text{c}}-1} \mathbf{x}_{n,m}^H \mathbf{A}\mathbf{x}_{n,m} e^{-\jmath 2 \pi l\frac{n}{N_{\text{c}}}} e^{\jmath 2\pi \nu \frac{m}{N_{\text{s}}}}                                                               \\
         & =\! \sum_{m = 0}^{N_{\text{s}}-1} \mathbf{x}_m^H \widetilde{\mathbf{A}}   \mathbf{D}^{\ast}_{l} \widetilde{\mathbf{A}}^H \mathbf{x}_m  e^{\jmath 2\pi \nu \frac{m}{N_{\text{s}}}} \\
         & = \mathbf{x}^H \widetilde{\mathbf{A}} \big( \mathbf{D}_{\nu} \otimes \mathbf{D}^{\ast}_{l} \big) \widetilde{\mathbf{A}}^H \mathbf{x},
    \end{align}
\end{subequations}
where we define
\begin{subequations}\label{eq:defination_Matrix}
    \begin{align}
        \mathbf{A}     & \triangleq \mathbf{a}_{\text{T}}(\theta_0) \mathbf{a}_{\text{T}}^H(\theta_0) , ~~~~~ \widetilde{\mathbf{A}} \triangleq \mathbf{I}_{N_{\text{s}}N_{\text{c}}} \otimes \mathbf{a}_{\text{T}}(\theta_0), \label{eq:A_tilde_def}\\
        \mathbf{x}_{m} & \triangleq \big[ \mathbf{x}_{0,m}^T, \mathbf{x}_{1,m}^T, \dots, \mathbf{x}^T_{N_{\text{c}}-1,m} \big]^T \in  \mathbb{C}^{N_{\text{c}}N_{\text{t}}},                       \\
        \mathbf{x}     & \triangleq \big[ \mathbf{x}^T_0, \mathbf{x}^T_1, \dots, \mathbf{x}^T_{N_{\text{s}}-1} \big]^T \in \mathbb{C}^{N_{\text{s}}N_{\text{c}}N_{\text{t}}}.
    \end{align}
\end{subequations}

Ideally, the ambiguity function should have a narrow mainlobe peak at  $(l = 0, \nu=0)$ and low sidelobes for $(l\ne0, \nu \ne 0)$ in order to provide good radar sensing performance. While deterministic waveforms such as chirp signals or ZC sequences often used in radar systems can provide excellent low-sidelobe performance, the dual-functional waveform in ISAC must also convey information, and thus will have a strong random component. Unless it is properly controlled, this random signal component inevitably exhibits higher sidelobe levels. Thus, the most important consideration in ISAC waveform design is how to control the sidelobes due to the embedded communication data, to eliminate ``ghost'' target peaks and to enable the detection of weak targets that would otherwise be disappear under the sidelobes. To quantify the range-Doppler sidelobe level of the ambiguity function, ISL is the most commonly used metric, and is defined as
\begin{subequations} \label{eq:ISL1_b}
    \begin{align}
        \xi_{\text{ISL}} & \triangleq \sum_{\nu = 0}^{N_{\text{s}}-1} \sum_{l = 0}^{N_{\text{c}}-1} \big|\chi(l, \nu)\big|^2 - \big|\chi(0, 0)\big|^2                                                                                                                      \\
                         & = \sum_{\nu = 0}^{N_{\text{s}}-1} \sum_{l = 0}^{N_{\text{c}}-1} \Big| \mathbf{x}^H \widetilde{\mathbf{A}} \big( \mathbf{D}_{\nu} \otimes \mathbf{D}^{\ast}_{l} \big) \widetilde{\mathbf{A}}^H \mathbf{x} \Big|^2 \notag \\
                         & \hspace{92pt} - |\mathbf{x}^H \widetilde{\mathbf{A}}  \widetilde{\mathbf{A}}^H \mathbf{x}|^2.
    \end{align}
\end{subequations}

We should emphasize that, in addition to range-Doppler sidelobes introduced by the transmit waveform, the received echo signal will also be distorted by noise. Both suppressing range-Doppler sidelobes and maintaining sufficient radar SNR are crucial for achieving superior sensing performance.
Nevertheless, measuring the radar receive SNR is challenging due to lack of information about the target RCS and noise power. Considering this reality in radar systems, the level of target illumination power can serve as an alternative metric, since increasing the transmission power towards the target will proportionally improve the receive SNR.  
Therefore, in addition to the ISL, we also incorporate consideration of the target illumination power in the waveform design to guarantee a satisfactory level of radar receive SNR. 
Specifically, we employ the transmit beampattern gain in the direction of the target to serve as a proxy for the receive SNR. For the time-domain transmit signal $\widetilde{\mathbf{x}}_{p, m}$, the total target illumination power during an OFDM frame can be written as
\begin{subequations}
    \begin{align}
        P_{\text{IL}}
         & = \sum_{m = 0}^{N_{\text{s}}-1} \sum_{p = 0}^{N_{\text{c}}-1}  \widetilde{\mathbf{x}}_{p,m}^H \mathbf{A} \widetilde{\mathbf{x}}_{p,m} \\
         & = \sum_{m = 0}^{N_{\text{s}}-1} \widetilde{\mathbf{x}}_{m}^H (\mathbf{I}_{N_{\text{c}}} \otimes \mathbf{A} ) \widetilde{\mathbf{x}}_{m} \\
         & = \widetilde{\mathbf{x}}^H (\mathbf{I}_{N_{\text{s}}N_{\text{c}}} \otimes \mathbf{A} ) \widetilde{\mathbf{x}} \\
         & = \mathbf{x}^H \widetilde{\mathbf{F}} (\mathbf{I}_{N_{\text{s}}N_{\text{c}}} \otimes \mathbf{A} ) \widetilde{\mathbf{F}}^H \mathbf{x}, \label{eq:PIL}
    \end{align}
\end{subequations}
where we define
\begin{subequations}
    \begin{align}
         & \!\!\! \widetilde{\mathbf{x}}_{m}   \triangleq  (\mathbf{F}_{N_\text{c}}^H \otimes \mathbf{I}_{N_{\text{t}}}) \, \mathbf{x}_{m} = \big[\,  \widetilde{\mathbf{x}}_{0,m}^T,  \dots, \,\widetilde{\mathbf{x}}^T_{N_{\text{c}}-1,m} \, \big]^T \\
         & \!\!\! \widetilde{\mathbf{F}}  \triangleq \mathbf{I}_{N_{\text{s}}} \!\otimes\! \mathbf{F}_{N_\text{c}} \!\otimes\! \mathbf{I}_{N_{\text{t}}}, ~ \widetilde{\mathbf{x}}  \triangleq \widetilde{\mathbf{F}}^H \! \mathbf{x} \!=\! \big[ \widetilde{\mathbf{x}}^T_0,  \dots, \widetilde{\mathbf{x}}^T_{N_{\text{s}}-1} \big]^T \!.
    \end{align}
\end{subequations}

\subsection{Communication Model and Performance Metric}
In addition to the radar sensing function, the BS uses the same dual-functional waveform, which is also to be designed to deliver information symbols to $K$ single-antenna users. We assume that the communication channels of the users experience frequency selective fading. The received signal at each user is down-converted to baseband, followed by analog-to-digital conversion, CP removal, and a DFT. Finally, the signal on the $n$-th subcarrier of the $k$-th user can be written as
\begin{equation} \label{eq:received signal}
    y_{n, m, k} = \mathbf{h}_{n, k}^{H} \mathbf{x}_{n,m} + z_{n, m, k},
\end{equation}
where $\mathbf{h}_{n, k} \in \mathbb{C}^{N_{\text{t}}}$ denotes the corresponding frequency-domain channel which is assumed to be known at the BS, and $z_{n, m, k} \sim \mathcal{CN} (0, \sigma_{\text{c}}^2)$ is AWGN.

In this paper, we propose the use of SLP rather than standard block-level precoding (BLP) in order to design the precoded signal $\mathbf{x}_{n, m}$. As explained below, the reason for doing so is to create additional DoFs for our waveform optimization that can be used for range-Doppler sidelobe suppression and handling the other ISAC constraints. To simplify the description of SLP, we assume that each modulated symbol $s_{n,m,k}$ is drawn from an $\Omega$-phase-shift-keying (PSK) constellation set $\mathcal{S}$, defined by
\begin{equation}
    s_{n,m,k} \in \mathcal{S} \triangleq \big\{ e^{\frac{\jmath \pi (2i-1)}{\Omega}}, i = 1, \dots,  \Omega \big\}.
\end{equation}
The waveform design can be extended to other types of constellations (e.g., higher-order QAM), but the required mathematics are more involved and PSK serves the purpose of illustrating the idea.

\begin{figure}[!t]
    \centering
    \includegraphics[width = 3 in]{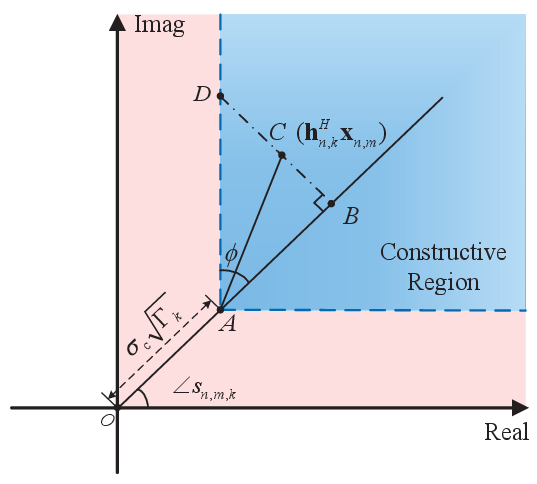}
    \caption{Constructive Interference Region for QPSK.}
    \label{fig:CR} \vspace{-2mm}
\end{figure}

Unlike standard linear BLP which attempts to eliminate the MUI, SLP instead converts the harmful MUI into CI using a nonlinear precoding design that provides additional DoFs that, in communication-only scenarios, can be used to reduce transmit power or increase robustness to noise and interference. In our application, we will use these DoFs to reduce the range-Doppler ISL.
To illustrate the concept of CI, we use quadrature-PSK (QPSK) as an example, and assume without loss of generality that $s_{n,m, k} = e^{\jmath\pi/4}$ is the transmit symbol. The CI concept is illustrated in Fig.~\ref{fig:CR}, where the received noise-free signal is $\mathbf{h}_{n, k}^{H} \mathbf{x}_{n,m}$, $\phi = \pi / \Omega$, the decision boundaries for the symbol $e^{\jmath\pi/4}$ are the positive halves of the $x$ and $y$ axes, and the constructive interference region for $s_{n,m, k}$ is the green sector. The transmitted symbol $s_{n,m, k}$ can be correctly detected if the received signal $y_{n, m, k}$ lies in the first quadrant. SLP designs the transmit waveform $\mathbf{x}_{n, m}$ to ensure that the noise-free received signal lies in the CI region away from the decision boundaries to provide some robustness to noise and interference.

To derive the mathematical formulation of the CI region, we project the noiseless received signal at point $C$ is projected onto the line in the direction of $\overrightarrow{OA}$ at point $B$. We define point $D$ as the intersection of CI region boundary with the extension of $\overrightarrow{BC}$. Then, for $\mathbf{h}_{n, k}^{H} \mathbf{x}_{n,m}$ to be in the CI region, the point $C$ should satisfy $|\overrightarrow{BD}|-|\overrightarrow{BC}| \ge 0$, which can be written as
\begin{equation}\label{eq:communication constraint}
    \begin{aligned}
         & \mathfrak{R}\big\{ \mathbf{h}_{n, k}^{H} \mathbf{x}_{n,m} s_{n,m,k}^{\ast} - \sigma_{\text{c}} \sqrt{\Gamma_{k}} \big\}  \tan \phi \\
         & ~~~~~~~ -\Big|\mathfrak{I}\big\{ \mathbf{h}_{n, k}^{H} \mathbf{x}_{n,m} s_{n,m,k}^{\ast} \big\}\Big| \ge 0, ~~ \forall k, n, m
    \end{aligned}
\end{equation}
where $\Gamma_{k}$ defines the ``buffer'' zone or safety margin between the CI region and the decision boundaries. The value of $\Gamma_k$ can be chosen to guarantee a certain minimum QoS for user $k$ in terms of SNR or symbol error rate. In order to reformulate (\ref{eq:communication constraint}) in a concise form, we define
\begin{subequations}
    \begin{align}
        \widetilde{\mathbf{h}}_{n,m, 2k}^H   & \triangleq \mathbf{h}_{n, k}^H s_{n,m,k}^{\ast} ( \sin\phi+ e^{-\jmath\frac{\pi}{2}}\cos\phi )  \\
        \widetilde{\mathbf{h}}_{n,m, 2k-1}^H & \triangleq \mathbf{h}_{n, k}^H s_{n,m,k}^{\ast}(\sin\phi- e^{-\jmath\frac{\pi}{2}}\cos\phi )    \\
        \gamma_{2k}                          & = \gamma_{2k-1} \triangleq \sigma \sqrt{\Gamma_{k}} \sin \phi, ~\forall  k.
    \end{align}
\end{subequations}
Then, the CI-based communication QoS constraints can be equivalently reformulated as
\begin{equation}\label{eq:comm constraint}
    \mathfrak{R}\big\{ \widetilde{\mathbf{h}}^H_{n,m,k'} \mathbf{x}_{n,m} \big\} \ge \gamma_{k'}, ~~ \forall k' = 1, 2, \dots, 2K, \forall n, m.
\end{equation}
In the remainder of this paper, we will employ (\ref{eq:comm constraint}) as the multi-user communication performance metric.

\newcounter{TempEqCnt}
\setcounter{TempEqCnt}{\value{equation}}
\setcounter{equation}{27}
\begin{figure*}[!t]
    \begin{subequations}\label{eq:ISL_der1}
        \begin{align}
            \xi_{\text{ISL}}
             & = \sum_{\nu = 0}^{N_{\text{s}}-1} \sum_{l = 0}^{N_{\text{c}}-1} \Big| \mathbf{x}^H \widetilde{\mathbf{A}} \big( \mathbf{D}_{\nu} \otimes \mathbf{D}^{\ast}_{l} \big) \widetilde{\mathbf{A}}^H \mathbf{x} \Big|^2 - |\mathbf{x}^H \widetilde{\mathbf{A}}  \widetilde{\mathbf{A}}^H \mathbf{x}|^2                                                                                                                                            \\
             & = \sum_{\nu = 0}^{N_{\text{s}}-1} \sum_{l = 0}^{N_{\text{c}}-1} \text{Tr} \Big\{ \big( \mathbf{D}^{\ast}_{\nu} \otimes \mathbf{D}_{l} \big) \widetilde{\mathbf{A}}^H \mathbf{x}  \mathbf{x}^H \widetilde{\mathbf{A}} \big(\mathbf{D}_{\nu} \otimes \mathbf{D}^{\ast}_{l} \big) \widetilde{\mathbf{A}}^H \mathbf{x} \mathbf{x}^H \widetilde{\mathbf{A}} \Big\} - |\mathbf{x}^H \widetilde{\mathbf{A}}  \widetilde{\mathbf{A}}^H \mathbf{x}|^2 \\
             & = \text{vec}^H(\widetilde{\mathbf{A}}^H \mathbf{x} \mathbf{x}^H \widetilde{\mathbf{A}}) \Big( \! \sum_{\nu = 0}^{N_{\text{s}}-1} \! \sum_{l = 0}^{N_{\text{c}}-1} \! \big( \mathbf{D}_{\nu} \otimes \mathbf{D}^{\ast}_{l}  \otimes  \mathbf{D}^{\ast}_{\nu} \otimes \mathbf{D}_{l} \big) - \mathbf{I} \Big)  \text{vec}(\widetilde{\mathbf{A}}^H \mathbf{x} \mathbf{x}^H \widetilde{\mathbf{A}})                                             \\
             & = \text{vec}^H(\widetilde{\mathbf{A}}^H \mathbf{x} \mathbf{x}^H \widetilde{\mathbf{A}}) \mathbf{B}  \text{vec}(\widetilde{\mathbf{A}}^H \mathbf{x} \mathbf{x}^H \widetilde{\mathbf{A}}).
        \end{align}
    \end{subequations}
    \rule[-0pt]{18.1 cm}{0.05em}
\end{figure*}
\setcounter{equation}{\value{TempEqCnt}}

\subsection{Problem Formulation}
Based on the above descriptions, we employ SLP to design the transmit waveform with the aim of minimizing the range-Doppler ISL while satisfying the illumination power constraint, the multi-user communication QoS requirements, and the transmit power budget. In addition, we also constrain the transmitted signal to be constant-modulus in order to avoid the non-linear distortion of the power amplifier and enhance its efficiency. For SLP, this requirement means that each element of the time-domain transmit vector $\widetilde{\mathbf{x}}$ should have a constant amplitude, and it can be mathematically expressed as
\begin{equation} \label{eq:constant_modulus1}
    \big| \widetilde{\mathbf{x}} \big| = \big| \widetilde{\mathbf{F}}^H \mathbf{x} \big| = \sqrt{P_{\text{T}}/ N_{\text{tot}}} \mathbf{1}_{N_{\text{tot}}},
\end{equation}
where $| \cdot |$ denotes an element-wise absolute value operation, $P_{\text{T}}$ is the transmit power allocated for an OFDM frame, and $N_{\text{tot}}= N_{\text{s}}  N_{\text{c}} N_{\text{t}}$. 

Thus, the overall SLP-based ISAC waveform design problem for range-Doppler sidelobe suppression can be formulated as
\begin{subequations}	\label{pro1}
    \begin{align}
         & \underset{\mathbf{x}}{\min} ~~~ \xi_{\text{ISL}}  \label{eq:pro1_ob}                                                                                                                                                                          \\
         & ~ \text{s.t.} \hspace{10pt}  \mathbf{x}^H \widetilde{\mathbf{F}} (\mathbf{I}_{N_{\text{s}}N_{\text{c}}} \otimes \mathbf{A} ) \widetilde{\mathbf{F}}^H \mathbf{x} \ge P_{\text{0}} , \label{eq:pro1_ILpower} \\
         & \hspace{27pt} \mathfrak{R}\big\{ \widetilde{\mathbf{h}}^H_{n,m,k'} \mathbf{x}_{n,m} \big\} \ge \gamma_{k'}, ~~ \forall k', n, m,                                                                                                                 \\
         & \hspace{27pt} \big| \widetilde{\mathbf{F}}^H \mathbf{x} \big| = \sqrt{P_{\text{T}}/ N_{\text{tot}}} \mathbf{1}_{N_{\text{tot}}}, \label{eq:pro1_cmcon}
    \end{align}
\end{subequations}
where $P_0$ is the required minimum target illumination power.
The proposed waveform design problem in~(\ref{pro1}) is challenging to solve due to the non-convex quartic objective function and the non-convex constraints in~(\ref{eq:pro1_ILpower}) and~(\ref{eq:pro1_cmcon}). In the next section, we develop an efficient algorithm to tackle these difficulties.

\section{SLP-based ISAC Waveform Design For Range-Doppler Sidelobe Suppression} \label{sec:algo}
In this section, we propose an MM-ADMM-based algorithm to solve the non-convex waveform design problem (\ref{pro1}). Specifically, we first reformulate the illumination power constraint (\ref{eq:pro1_ILpower}) into a convex form. Then, we employ the MM method \cite{MM} to convert (\ref{pro1}) into a series of more manageable sub-problems and utilize the ADMM method \cite{ADMM} to handle the constant-modulus constraint (\ref{eq:pro1_cmcon}). The algorithm then alternates between solving each sub-problem. The details of the algorithm development are described below.

\subsection{Problem Reformulation}

We first equivalently reformulate the illumination power constraint in~(\ref{eq:PIL}) as
\setcounter{equation}{28}
\begin{subequations}\label{eq:PIL reform}
    \begin{align}
        P_{\text{IL}}
        = & \mathbf{x}^H \widetilde{\mathbf{F}} (\mathbf{I}_{N_{\text{s}}N_{\text{c}}} \otimes \mathbf{A} ) \widetilde{\mathbf{F}}^H \mathbf{x} - N_{\text{t}}P_{\text{T}} + N_{\text{t}}P_{\text{T}}                          \\
        = & \mathbf{x}^H \widetilde{\mathbf{F}} (\mathbf{I}_{N_{\text{s}}N_{\text{c}}} \otimes \mathbf{A} ) \widetilde{\mathbf{F}}^H \mathbf{x} - N_{\text{t}} \mathbf{x}^H \mathbf{x} + N_{\text{t}}P_{\text{T}}              \\
        = & - \mathbf{x}^H \widetilde{\mathbf{F}} \big(\mathbf{I}_{N_{\text{s}}N_{\text{c}}} \otimes (N_{\text{t}}\mathbf{I}_{N_{\text{t}}} - \mathbf{A}) \big) \widetilde{\mathbf{F}}^H \mathbf{x} + N_{\text{t}}P_{\text{T}} \\
        = & - \mathbf{x}^H \overline{\mathbf{A}} \mathbf{x} + N_{\text{t}}P_{\text{T}},  \label{eq:IL_reformulate}
    \end{align}
\end{subequations}
where we have used the power constraint $\mathbf{x}^H\mathbf{x} = P_\text{T}$ derived from the constant-modulus constraint in (\ref{eq:PIL reform}b) and the fact that 
$\widetilde{\mathbf{F}}\widetilde{\mathbf{F}}^H=\mathbf{I}_{N_{\text{tot}}}$ for the DFT matrix in (\ref{eq:PIL reform}c), and we define
\begin{equation}
    \overline{\mathbf{A}}  \triangleq \widetilde{\mathbf{F}} \big( \mathbf{I}_{N_{\text{s}}N_{\text{c}}} \otimes (N_{\text{t}}\mathbf{I}_{N_{\text{t}}} - \mathbf{A}) \big) \widetilde{\mathbf{F}}^H.
\end{equation}
Recalling the matrix definition in (\ref{eq:A_tilde_def}), the maximum eigenvalue of $\mathbf{A}$ is $N_{\text{t}}$. Thus, the eigenvalues of matrix $N_{\text{t}}\mathbf{I}_{N_{\text{t}}}-\mathbf{A}$ are greater than or equal to 0, making it a positive semi-definite matrix. According to the properties of the Kronecker product and the DFT matrix, it is evident that $\overline{\mathbf{A}}$ is also a positive semi-definite. Substituting (\ref{eq:IL_reformulate}) into (\ref{eq:pro1_ILpower}), the illumination power constraint can be converted into a convex form: 
\begin{equation}
    \mathbf{x}^H \overline{\mathbf{A}} \mathbf{x} \leq  \bar{P}_0,
\end{equation}
where $\bar{P}_0 = N_{\text{t}}P_{\text{T}}-P_{0}$.

To facilitate the subsequent algorithm development, we reformulate the range-Doppler ISL expression in~(\ref{eq:ISL1_b}) into a more compact form with respect to the transmit waveform $\mathbf{x}$. After some algebraic manipulations presented in (\ref{eq:ISL_der1}) at the top of this page, the range-Doppler ISL can be equivalently written as
\begin{equation} 
    \xi_{\text{ISL}} = \text{vec}^H(\widetilde{\mathbf{A}}^H \mathbf{x} \mathbf{x}^H \widetilde{\mathbf{A}}) \mathbf{B} \text{vec}(\widetilde{\mathbf{A}}^H \mathbf{x} \mathbf{x}^H \widetilde{\mathbf{A}}),
\end{equation}
where we define
\begin{equation} \label{eq:defination_B}
    \mathbf{B} \triangleq \sum_{\nu = 0}^{N_{\text{s}}-1} \sum_{l = 0}^{N_{\text{c}}-1} \big( \mathbf{D}_{\nu} \otimes \mathbf{D}^{\ast}_{l}  \otimes  \mathbf{D}^{\ast}_{\nu} \otimes \mathbf{D}_{l} \big) - \mathbf{I}_{N^2_{\text{s}} N^2_{\text{c}}}.
\end{equation}
It is clear that $\mathbf{B}$ is diagonal since $\mathbf{D}_\nu$ and $\mathbf{D}_l$ are both diagonal. Moreover, by using the properties of the Kronecker product and the matrix definitions in (\ref{eq:D_l_define}), (\ref{eq:D_nu_define}), and (\ref{eq:defination_B}), we can further calculate each element of $\mathbf{B}$ as follows. Recalling the matrix definitions $\mathbf{D}_{l}(n+1, n+1) = e^{\jmath 2 \pi l \frac{n}{N_{\text{c}}}}$ and $\mathbf{D}_{\nu}(m+1, m+1) = e^{\jmath 2 \pi \nu \frac{m}{N_{\text{s}}} }$, the $(j+1)$-th diagonal element of $\mathbf{B}$ can be expressed as
\begin{equation}\label{eq:Bj}
    \mathbf{B}(j+1,j+1) = \sum_{l = 0}^{N_{\text{c}}-1} e^{\jmath 2 \pi l\frac{n-n'}{N_{\text{c}}}} \sum_{\nu = 0}^{N_{\text{s}}-1} e^{\jmath 2 \pi \nu \frac{m-m'}{N_{\text{s}}}} -1,
\end{equation}
where $n= \text{mod}(j, N_{\text{c}})$, $m'=\text{mod}(\lfloor {j}/{N_{\text{c}}} \rfloor,N_{\text{s}})$, $n'=\text{mod}(\lfloor {j}/N_{\text{s}}/N_{\text{c}} \rfloor,N_{\text{c}})$, and $m=\text{mod}(\lfloor {j}/N_{\text{s}}/N^2_{\text{c}} \rfloor,N_{\text{s}})$. 
We observe that the first term on the right-hand side of (\ref{eq:Bj}) equals $N_\text{s}N_\text{c}$ if both $(n-n')/N_{\text{c}}$ and $(m-m')/{N_{\text{s}}}$ are integers; otherwise it will be 0.
Thus, by defining $\theta_1 \triangleq (n-n')/N_{\text{c}}$ and $\theta_2 \triangleq (m-m')/{N_{\text{s}}}$, the $(j+1)$-th diagonal element of $\mathbf{B}$ can be further written as
\begin{equation} \label{eq:B_deri}
    \mathbf{B}(j+1,j+1) =
    \begin{cases}
        N_{\text{s}} N_{\text{c}}-1, ~~ & \theta_1, \theta_2 \in \mathbb{Z} \\
        -1,                         ~~  & \text{otherwise}.
    \end{cases}
\end{equation}
Therefore, the original problem in~(\ref{pro1}) can be equivalently reformulated as
\begin{subequations}	\label{proCM_re}
    \begin{align}
         & \underset{\mathbf{x}}{\min} ~~~  \text{vec}^H(\widetilde{\mathbf{A}}^H \mathbf{x} \mathbf{x}^H \widetilde{\mathbf{A}}) \mathbf{B} \text{vec}(\widetilde{\mathbf{A}}^H \mathbf{x} \mathbf{x}^H \widetilde{\mathbf{A}})  \label{eq:proCM_re_ob} \\
         & ~ \text{s.t.} \hspace{12pt}  \mathbf{x}^H \overline{\mathbf{A}} \mathbf{x} \leq \bar{P}_0  , \label{eq:proCM_re_ILpower}                                                                                                                            \\
         & \hspace{27pt} \mathfrak{R}\big\{ \widetilde{\mathbf{h}}^H_{n,m,k'} \mathbf{x}_{n,m} \big\} \ge \gamma_{k'}, ~~ \forall k', n, m,                                                                                                                 \\
         & \hspace{27pt} \big| \widetilde{\mathbf{F}}^H \mathbf{x} \big| = \sqrt{\frac{P_{\text{T}}}{N_{\text{tot}}}} \mathbf{1}_{N_{\text{tot}}}.  \label{eq:proCM_rewaveform_con}
    \end{align}
\end{subequations}

\subsection{MM Transformation}
To efficiently solve the non-convex waveform design problem in~(\ref{proCM_re}), we utilize the MM method \cite{MM} to transform the original problem into a series of more tractable sub-problems. In particular, we seek a surrogate upper bound that locally approximates the objective function in~(\ref{eq:proCM_re_ob}), and then we minimize the surrogate function in each iteration. The procedure for deriving the surrogate function is described next.

An upper-bound surrogate for a general quadratic form $\mathbf{x} ^H \mathbf{L} \mathbf{x}$ can be derived via a second-order Taylor expansion series at the current point $\mathbf{x}_t$ as:
\setcounter{equation}{37}
\begin{equation} \label{eq:MM}
    \begin{aligned}
         & \mathbf{x} ^H \mathbf{L} \mathbf{x} \leq \lambda_{\text{L}} \mathbf{x}^H  \mathbf{x}  + 2 \mathfrak{R} \big\{ \mathbf{x}_t ^H (\mathbf{L} - \lambda_{\text{L}} \mathbf{I}) \mathbf{x} \big\} \\
         & \hspace{2.5cm}+  \mathbf{x}_t ^H (\lambda_{\text{L}} \mathbf{I} - \mathbf{L}) \mathbf{x}_t,
    \end{aligned}
\end{equation}
where $\mathbf{L}$ is a Hermitian matrix, and $\lambda_{\text{L}}$ is the largest eigenvalue of $\mathbf{L}$. 
Based on (\ref{eq:MM}), an surrogate upper-bound for the range-Doppler ISL at point $\mathbf{x}_t$ can be constructed as
\begin{subequations}
    \begin{align}
        \! \xi_{\text{ISL}}
         & = \text{vec}^H(\widetilde{\mathbf{A}}^H \mathbf{x} \mathbf{x}^H \widetilde{\mathbf{A}}) \mathbf{B} \text{vec}(\widetilde{\mathbf{A}}^H \mathbf{x} \mathbf{x}^H \widetilde{\mathbf{A}}) \\
         & \leq  \lambda_{\text{B}} f_1(\mathbf{x}) + f_2(\mathbf{x}) + c_1, \label{eq:ISL_majorize_result}
    \end{align}
\end{subequations}
where $\lambda_{\text{B}} = N_{\text{s}} N_{\text{c}} -1$ according to the definition of $\mathbf{B}$ in~(\ref{eq:B_deri}), and for brevity we define
\begin{subequations}
    \begin{align}
        f_1(\mathbf{x}) & \triangleq \text{vec}^H(\widetilde{\mathbf{A}}^H \mathbf{x} \mathbf{x}^H \widetilde{\mathbf{A}}) \text{vec}(\widetilde{\mathbf{A}}^H \mathbf{x} \mathbf{x}^H \widetilde{\mathbf{A}}) \\
        f_2(\mathbf{x}) & \triangleq 2 \mathfrak{R} \! \big\{ \! \text{vec}^H \! (\widetilde{\mathbf{A}}\!^H \!\mathbf{x}_t \mathbf{x}_t^H \! \widetilde{\mathbf{A}}) (\mathbf{B}\!-\!\lambda_{\text{B}}\mathbf{I}_{N^2_{\text{s}}N^2_{\text{c}}}\!) \text{vec}(\widetilde{\mathbf{A}}\!^H \!\mathbf{x} \mathbf{x}^H \! \widetilde{\mathbf{A}}) \! \big\} \\
        c_1             & \triangleq  \text{vec}^H(\widetilde{\mathbf{A}}^H \mathbf{x}_t \mathbf{x}_t^H \widetilde{\mathbf{A}})  (\lambda_{\text{B}} \mathbf{I}_{N^2_{\text{s}} N^2_{\text{c}}}-\mathbf{B}) \text{vec}(\widetilde{\mathbf{A}}^H \mathbf{x}_t \mathbf{x}_t^H \widetilde{\mathbf{A}}).
    \end{align}
\end{subequations}
It can be seen that $f_1(\mathbf{x})$ is a quartic function w.r.t. $\mathbf{x}$, $f_2(\mathbf{x})$ is a quadratic function w.r.t. $\mathbf{x}$, and $c_1$ is a constant irrelevant to the optimizing variable $\mathbf{x}$. 
In order to construct a favorable convex upper-bound for $\xi_{\text{ISL}}$, we further propose to majorize $f_1(\mathbf{x})$ and $f_2(\mathbf{x})$ again as follows.

\newcounter{TempEqCnt2}
\setcounter{TempEqCnt2}{\value{equation}}
\setcounter{equation}{36}
\begin{figure*}[!t]
    \begin{subequations}\label{eq:f1_der1}
        \begin{align}
            f_1(\mathbf{x})
             & = \text{vec}^H({\mathbf{x}\mathbf{x}^H}) \big( (\widetilde{\mathbf{A}} \widetilde{\mathbf{A}}^H)^T \otimes (\widetilde{\mathbf{A}} \widetilde{\mathbf{A}}^H) \big) \text{vec}({\mathbf{x}\mathbf{x}^H}) \\
             & \leq \lambda_{\text{C}} \text{vec}^H({\mathbf{x}\mathbf{x}^H}) \text{vec}({\mathbf{x}\mathbf{x}^H}) \!+\! 2\mathfrak{R}\big\{ \text{vec}^H({\mathbf{x}_t\mathbf{x}_t^H})  (\mathbf{C} \!-\! \lambda_{\text{C}} \mathbf{I}_{N_{\text{tot}}^2} )  \text{vec}({\mathbf{x}\mathbf{x}^H}) \}   \!+\! \text{vec}^H({\mathbf{x}_t \mathbf{x}^H_t})  (\lambda_{\text{C}} \mathbf{I}_{N_{\text{tot}}^2} - \mathbf{C} ) \text{vec}({\mathbf{x}_t \mathbf{x}_t^H})    \\
             & = N_{\text{t}}^2 P_{\text{T}}^2 + \mathbf{x}^H \mathbf{G}_t \mathbf{x} +  \text{vec}^H({\mathbf{x}_t \mathbf{x}^H_t})  (N_{\text{t}}^2 \mathbf{I}_{N_{\text{tot}}^2} \!-\! \mathbf{C} ) \text{vec}({\mathbf{x}_t \mathbf{x}_t^H})  \label{eq:f1_stepc}   \\
             & \leq  N_{\text{t}}^2 P_{\text{T}}^2 + \lambda_{\text{G}_t} \mathbf{x}^H \mathbf{x} \!+\! 2 \mathfrak{R} \big\{ \mathbf{x}_t^H (\mathbf{G}_t \!-\! \lambda_{\text{G}_t}\mathbf{I}_{N_{\text{tot}}}) \mathbf{x} \big\} \!+\! \mathbf{x}_t^H (\lambda_{\text{G}_t}\mathbf{I}_{N_{\text{tot}}} \!-\! \mathbf{G}_t) \mathbf{x}_t + \text{vec}^H({\mathbf{x}_t \mathbf{x}^H_t})  (N_{\text{t}}^2 \mathbf{I}_{N_{\text{tot}}^2} - \mathbf{C} ) \text{vec}({\mathbf{x}_t \mathbf{x}_t^H}) \! \\
             & = N_{\text{t}}^2 P_{\text{T}}^2 + \lambda_{\text{G}_t} P_{\text{T}} + 2 \mathfrak{R} \big\{ \mathbf{x}_t^H (\mathbf{G}_t - \lambda_{\text{G}_t}\mathbf{I}_{N_{\text{tot}}}) \mathbf{x} \big\} +\!   \mathbf{x}_t^H (\lambda_{\text{G}_t}\mathbf{I}_{N_{\text{tot}}} \!-\! \mathbf{G}_t) \mathbf{x}_t  + \text{vec}^H({\mathbf{x}_t \mathbf{x}^H_t})  (N_{\text{t}}^2 \mathbf{I}_{N_{\text{tot}}^2} - \mathbf{C} ) \text{vec}({\mathbf{x}_t \mathbf{x}_t^H})   \label{eq:f1_stepe}   \\
             & = 2 \mathfrak{R} \big\{ \mathbf{x}_t^H (\mathbf{G}_t - \lambda_{\text{G}_t}\mathbf{I}_{N_{\text{tot}}}) \mathbf{x} \big\} + c_2.
        \end{align}
    \end{subequations}
    \rule[-0pt]{18.1 cm}{0.05em}
\end{figure*}
\setcounter{equation}{\value{TempEqCnt2}}

The quartic function $f_1(\mathbf{x})$ can be majorized by a simple linear function using the second-order Taylor expansion in (\ref{eq:MM}) twice, which can lead to
\begin{equation} \label{eq:f1_majorize}
    f_1(\mathbf{x}) \leq 2 \mathfrak{R} \big\{ \mathbf{x}_t^H (\mathbf{G}_t - \lambda_{\text{G}_t} \mathbf{I}_{N_{\text{tot}}}) \mathbf{x} \big\} + c_2.
\end{equation}
The details of the derivation are presented in (\ref{eq:f1_der1}) at the top of this page, where we define
\begin{subequations}
    \begin{align}
        \mathbf{C}   & \triangleq (\widetilde{\mathbf{A}} \widetilde{\mathbf{A}}^H)^T \otimes (\widetilde{\mathbf{A}} \widetilde{\mathbf{A}}^H)                                                                                         \\
        \mathbf{G}_t & \triangleq  2 (\widetilde{\mathbf{A}} \widetilde{\mathbf{A}}^H \mathbf{x}_t \mathbf{x}_t^H \widetilde{\mathbf{A}} \widetilde{\mathbf{A}}^H - \lambda_{\text{C}} \mathbf{x}_t \mathbf{x}_t^H)                      \\
        c_2          & \triangleq \lambda_{\text{C}} P_{\text{T}}^2 + \lambda_{\text{G}_t} P_{\text{T}} + \mathbf{x}_t^H (\lambda_{\text{G}_t}\mathbf{I}_{N_{\text{tot}}} - \mathbf{G}_t) \mathbf{x}_t                             \notag \\
                     & \hspace{1.3cm}+ \text{vec}^H({\mathbf{x}_t \mathbf{x}^H_t})  (\lambda_{\text{C}} \mathbf{I}_{N_{\text{tot}}^2} - \mathbf{C} ) \text{vec}({\mathbf{x}_t \mathbf{x}_t^H}),
    \end{align}
\end{subequations}
and the scalars $\lambda_{\text{C}}$ and $\lambda_{\text{G}_t}$ are the largest eigenvalues of matrices $\mathbf{C}$ and $\mathbf{G}_t$, respectively.  
It is easy to show that the largest eigenvalue of matrix $\mathbf{C}$ is constant, i.e., $\lambda_{\text{C}} = N_{\text{t}}^2$, according to the definition of $\mathbf{C}$ and $\widetilde{\mathbf{A}}$ in (\ref{eq:A_tilde_def}). With the power constraint $\mathbf{x}^H\mathbf{x} = P_\text{T}$, (\ref{eq:f1_stepc}) is obtained using
\begin{equation}
    \text{vec}^H(\mathbf{x}\mathbf{x}^H) \text{vec}(\mathbf{x}\mathbf{x}^H) \! = \! P_{\text{T}}^2.
\end{equation}

Using the result in \eqref{eq:MM}, the quadratic function $f_2(\mathbf{x})$ can be majorized as follows:
\begin{subequations}
    \begin{align}
        f_2(\mathbf{x})
         & = \mathbf{x}^H \mathbf{M}_t \mathbf{x}  \\
         & \leq \lambda_{\text{M}_t} \mathbf{x}^H \mathbf{x} + 2 \mathfrak{R}\big\{ \mathbf{x}_t^H (\mathbf{M}_t - \lambda_{\text{M}_t} \mathbf{I}_{N_{\text{tot}}} ) \mathbf{x}\big\}  \notag \\
         & \hspace{49pt} +\mathbf{x}_t^H (\lambda_{\text{M}_t} \mathbf{I}_{N_{\text{tot}}} - \mathbf{M}_t ) \mathbf{x}_t \label{eq:f2_majorize_stepc}  \\
         & = 2\mathfrak{R} \big\{ \mathbf{x}_t^H (\mathbf{M}_t - \lambda_{\text{M}_t} \mathbf{I}_{N_{\text{tot}}}) \mathbf{x} \big\} + c_3 \label{eq:f2_majorize},
    \end{align}
\end{subequations}
where we define
\begin{subequations}
    \begin{align}
        \mathbf{M}_t
         & \triangleq \sum_{\nu=0}^{N_{\text{s}}-1} \! \sum_{l=0}^{N_{\text{c}}-1} \! \widetilde{\mathbf{A}}\big( \mathbf{D}_{\nu} \!\otimes\! \mathbf{D}^{\ast}_{l} \big)\widetilde{\mathbf{A}}\!^H \!\mathbf{x}_t \mathbf{x}_t^H \!\widetilde{\mathbf{A}} \big( \mathbf{D}_{\nu}^{\ast} \!\otimes\! \mathbf{D}_{l} \big) \widetilde{\mathbf{A}}^H \notag \\
         & \hspace{2cm}- (\lambda_{\text{B}} + 1)	\widetilde{\mathbf{A}} \widetilde{\mathbf{A}}^H \mathbf{x}_t \mathbf{x}_t^H   \widetilde{\mathbf{A}} \widetilde{\mathbf{A}}^H \\
        c_3
         & \triangleq            \lambda_{\text{M}_t} P_{\text{T}} + \mathbf{x}_t^H (\lambda_{\text{M}_t} \mathbf{I}_{N_{\text{tot}}}- \mathbf{M}_t ) \mathbf{x}_t,
    \end{align}
\end{subequations}
and $\lambda_{\text{M}_t}$ is the largest eigenvalue of $\mathbf{M}_{t}$. Since $\mathbf{D}_{\nu}$ and $\mathbf{D}_{l}$ are diagonal, $\mathbf{M}_t$ is Hermitian. This property enables the use of~(\ref{eq:MM}) to derive a surrogate upper-bound for $f_2(\mathbf{x})$ in (\ref{eq:f2_majorize_stepc}). In addition, we propose to employ the following definition to facilitate the algorithm development:
\begin{subequations}
    \begin{align}
        \mathbf{M}_t             & \triangleq \widetilde{\mathbf{A}} \widetilde{\mathbf{M}} \widetilde{\mathbf{A}}^H  \\
        \widetilde{\mathbf{M}}_t & \triangleq \text{mat}_{N_{\text{s}}N_{\text{c}} \times N_{\text{s}}N_{\text{c}}} \big\{  2(\mathbf{B}  -  \lambda_{\text{B}}\mathbf{I}_{N_{\text{s}}N_{\text{c}}}) \text{vec}(\widetilde{\mathbf{A}}^H \mathbf{x}_t \mathbf{x}_t^H \widetilde{\mathbf{A}} ) \big\}.
    \end{align}
\end{subequations}

Substituting the inequalities in (\ref{eq:f1_majorize}) and (\ref{eq:f2_majorize}) into (\ref{eq:ISL_majorize_result}), the surrogate upper-bound for the objective function $\xi_{\text{ISL}}$ can be written as
\begin{equation}    \label{eq:ISL_linear}
        \xi_{\text{ISL}} \leq \mathfrak{R} \big\{\mathbf{g}_t^H \mathbf{x} \big\} + c_4 ,
\end{equation}
where for brevity we define
\begin{subequations}
    \begin{align}
        \mathbf{g}_t & \triangleq 2 \lambda_{\text{B}} (\mathbf{G}_t - \lambda_{\text{G}_t} \mathbf{I}_{N_{\text{tot}}})\mathbf{x}_t + 2 (\mathbf{M}_t - \lambda_{\text{M}_t} \mathbf{I}_{N_{\text{tot}}})\mathbf{x}_t \\
        c_4          & \triangleq c_1+ \lambda_{\text{B}}c_2 + c_3,
    \end{align}
\end{subequations}
and where the constant term $c_4$ is irrelevant to the variable $\mathbf{x}$ and can be neglected.

Based on the above derivations, the waveform design problem around point $\mathbf{x}_t$ can be formulated as
\begin{subequations}	\label{proCM_MM}
    \begin{align}
         & \underset{\mathbf{x}}{\min} ~~~  \mathfrak{R} \big\{\mathbf{g}_t^H \mathbf{x} \big\}  \label{eq:proCM_MM_ob}                                                               \\
         & ~ \text{s.t.} \hspace{12pt}  \mathbf{x}^H \overline{\mathbf{A}} \mathbf{x} \leq \bar{P}_0 , \label{eq:proCM_MM_ILpower}                                                          \\
         & \hspace{27pt} \mathfrak{R}\big\{ \widetilde{\mathbf{h}}^H_{n,m,k'} \mathbf{x}_{n,m} \big\} \ge \gamma_{k'}, ~~ \forall k', n, m,                                              \\
         & \hspace{27pt} \big| \widetilde{\mathbf{F}}^H \mathbf{x} \big| = \sqrt{\frac{P_{\text{T}}}{N_{\text{tot}}}}  \mathbf{1}_{N_{\text{tot}}}.  \label{eq:proCM_MM_waveform_con}
    \end{align}
\end{subequations}
While the objective function (\ref{eq:proCM_MM_ob}) and the illumination power constraint (\ref{eq:proCM_MM_ILpower}) are convex sub-problem (\ref{proCM_MM}) remains non-convex due to the constant-modulus constraint (\ref{eq:proCM_MM_waveform_con}). Next, we employ the ADMM method to handle this constraint.

\subsection{ADMM Transformation}
To decouple the non-convex constraint (\ref{eq:proCM_MM_waveform_con}) and other convex constraints, a new auxiliary variable $\mathbf{z} \triangleq [z_1, \dots, z_{N_{\text{s}}N_{\text{c}}N_{\text{t}}}]$ and a corresponding equality constraint $\mathbf{z} = \widetilde{\mathbf{F}}^H \mathbf{x}$ are introduced as follows:
\begin{subequations}\label{eq:problem_dual_vari}
    \begin{align}
         & \underset{\mathbf{x},\mathbf{z}}{\min}~~~  \mathfrak{R} \big\{ \mathbf{g}_t^H \mathbf{x}\big\}                                                                 \\
         & ~\text{s.t.} \hspace{12pt}  \mathbf{x}^H \overline{\mathbf{A}} \mathbf{x} \leq \bar{P}_0  , \label{eq:cons_1_x}                                                      \\
         & \hspace{27pt} \mathfrak{R}\big\{ \widetilde{\mathbf{h}}^H_{n,m,k'} \mathbf{x}_{n,m} \big\} \ge \gamma_{k'}, ~~ \forall k', n, m,   \label{eq:cons_2_x}            \\
         & \hspace{27pt}\big| \widetilde{\mathbf{F}}^H \mathbf{x} \big| \leq \sqrt{\frac{P_{\text{T}}}{N_{\text{tot}}}}  \mathbf{1}_{N_{\text{tot}}}, \label{eq:cons_3_x} \\
         & \hspace{27pt} \widetilde{\mathbf{F}}^H \mathbf{x} = \mathbf{z},                                                                                                \\
         & \hspace{27pt} | \mathbf{z} | = \sqrt{\frac{P_{\text{T}}}{N_{\text{tot}}}}  \mathbf{1}_{N_{\text{tot}}}.
    \end{align}
\end{subequations}
In order to employ the ADMM framework, we define the set $\mathcal{X}$ as the feasible region encompassing the inequality constraints (\ref{eq:cons_1_x}), (\ref{eq:cons_2_x}), and (\ref{eq:cons_3_x}), and the corresponding indicator function $\mathbb{I}_{\mathcal{X}} (\mathbf{x})$ as
            \begin{equation}
                \mathbb{I}_{\mathcal{X}} (\mathbf{x})  \triangleq
                \begin{cases}
                    0        & ~~\mathbf{x} \in \mathcal{X}; \\
                    + \infty & ~~\text{otherwise}.
                \end{cases}
            \end{equation}
Incorporating the feasibility indicator function in the objective function, 
problem~(\ref{eq:problem_dual_vari}) becomes
\begin{subequations}\label{eq:problem_admm_1}
    \begin{align}
         & \underset{\mathbf{x},\mathbf{z}}{\min}~~~  \mathfrak{R} \big\{ \mathbf{g}_t^H \mathbf{x} \big\} + \mathbb{I}_{\mathcal{X}}(\mathbf{x}) \\
         & ~\text{s.t.}\hspace{12pt}  \widetilde{\mathbf{F}}^H \mathbf{x} = \mathbf{z},                                                           \\
         & \hspace{27pt} | \mathbf{z}| = \sqrt{\frac{P_{\text{T}}}{N_{\text{tot}}}}  \mathbf{1}_{N_{\text{tot}}}.
    \end{align}
\end{subequations}
Problem (\ref{eq:problem_admm_1}) can be solved by minimizing its augmented Lagrangian function, given by
\begin{equation}\label{eq:AL_fun}
    \begin{aligned}
        \!\!\!\! \mathcal{L}_1(\mathbf{x}, \mathbf{z}, \boldsymbol{\lambda}, \boldsymbol{\mu}) & \! \triangleq \mathfrak{R} \big\{ \mathbf{g}_t^H \mathbf{x} \big\} \! + \! \mathbb{I}_{\mathcal{X}}(\mathbf{x}) \!+\! \frac{\rho}{2} \big\| \widetilde{\mathbf{F}}^H \mathbf{x} \!-\! \mathbf{z} + \frac{\boldsymbol{\lambda}}{\rho} \big\|^2 \\
                         & \hspace{1.2cm} + \frac{\rho}{2} \big\| |\mathbf{z}| -  \sqrt{\frac{P_{\text{T}}}{N_{\text{tot}}}} \mathbf{1}_{N_{\text{tot}}}  + \frac{\boldsymbol{\mu}}{\rho}  \big\|^2,
    \end{aligned}
\end{equation}
where $\boldsymbol{\lambda} \in \mathbb{C}^{N_{\text{tot}}}$ and $\boldsymbol{\mu} \in \mathbb{C}^{N_{\text{tot}}}$ are the dual variables and $\rho > 0 $ is a penalty parameter. The augmented Lagrangian can be minimized by alternately updating $\mathbf{x}$, $\mathbf{z}$, $\boldsymbol{\lambda}$ and $\boldsymbol{\mu}$.

\begin{algorithm}[!t]
    \begin{small}
        \caption{Proposed MM-ADMM Algorithm}
        \label{alg:1}
        \begin{algorithmic}[1]
            \REQUIRE {$\widetilde{\mathbf{h}}^H_{n,m,k'}$, $\gamma_{k'}$, $\forall n, m, k'$, $\mathbf{a}_{\text{T}}(\theta_0)$, $\mathbf{B}$, $\widetilde{\mathbf{F}}$,  $\rho$, $P_{\text{T}}$, $\bar{P}_0$, $\delta_{\text{th}}$.}
            \ENSURE {$\mathbf{x}^\star$.}
            \STATE {Initialize $\mathbf{x}_0$, $t:=0$.}
            \STATE {Calculate the objective value $\xi_{\text{ISL}}$ using (\ref{eq:proCM_re_ob}).}
            \REPEAT
            \STATE {$\hat{\xi}_{\text{ISL}}:=\xi_{\text{ISL}}$.}
            \STATE {Calculate $\lambda_{\text{G}_t}$, $\lambda_{\text{M}_t}$, and $\mathbf{g}_t$.}
            \STATE {Initialize $u:=0$, $\mathbf{x}_t^{0} \!:= \mathbf{x}_t$, $\mathbf{z}^{0} \!:= \widetilde{\mathbf{F}}^H \mathbf{x}_t^{0}$\!, $\boldsymbol{\lambda}^{0}\!:=\mathbf{0}$, $\boldsymbol{\mu}^{0}\!:=\mathbf{0}$.}
            \REPEAT
            \STATE {Update $\mathbf{x}_{t}^{u+1}$ by solving (\ref{eq:pro_update_x}).}
            \STATE Update {$\mathbf{z}^{u+1}$ via (\ref{eq:up_z}).}
            \STATE Update {$\boldsymbol{\lambda}^{u+1}$ and $\boldsymbol{\mu}^{u+1}$ using (\ref{eq:up_dual_vari}).}
            \STATE {$u:=u+1$.}
            \UNTIL {$\| \widetilde{\mathbf{F}}^H \! \mathbf{x}_t^{u} \!- \mathbf{z}^{u} \|^2 \leq \delta_{\text{th}}$ $\&$ $ \| |\mathbf{z}^{u}| \!- \!\sqrt{P_{\text{T}}/N_{\text{tot}}} \mathbf{1}_{N_{\text{tot}}} \|^2 \leq \delta_{\text{th}} $.}
            \STATE {$\mathbf{x}_{t+1}:= \mathbf{x}_t^{u}$.}
            \STATE {Calculate $\xi_{\text{ISL}}$, $t:=t+1$.}
            \UNTIL {$|\xi_{\text{ISL}} - \hat{\xi}_{\text{ISL}} | / \hat{\xi}_{\text{ISL}} \leq \delta_{\text{th}}$}
            \STATE {Return $\mathbf{x}^\star = \mathbf{x}_t$.}
        \end{algorithmic}
    \end{small}
\end{algorithm}

\subsection{Block Update}
\textit{1) Update} $\mathbf{x}$: For fixed $\mathbf{z}$, $\boldsymbol{\lambda}$ and $\boldsymbol{\mu}$, the update for $\mathbf{x}$ can be obtained by solving the following problem:
\begin{equation}
    \underset{\mathbf{x}}{\min}~~~ \mathfrak{R} \big\{ \mathbf{g}_t^H \mathbf{x} \big\} + \mathbb{I}_{\mathcal{X}}(\mathbf{x}) + \frac{\rho}{2} \big\| \widetilde{\mathbf{F}}^H \mathbf{x} - \mathbf{z} + \rho^{-1} \boldsymbol{\lambda} \big\|^2.
\end{equation}
To facilitate the algorithm development, we omit the terms in the objective function that are irrelevant to $\mathbf{x}$ and incorporate the indicator function $\mathbb{I}_{\mathcal{X}}(\mathbf{x})$ into the constraints, resulting in an equivalent concise form of the original optimization problem:
\begin{subequations}\label{eq:pro_update_x}
    \begin{align}
         & \underset{\mathbf{x}}{\min}~~~ \mathfrak{R} \big\{ \mathbf{m}_t^H \mathbf{x}  \big\} + \frac{\rho}{2} \big\|  \mathbf{x} \big\|^2          \\
         & ~ \text{s.t.} \hspace{12pt}  \mathbf{x}^H \overline{\mathbf{A}} \mathbf{x} \leq \bar{P}_0  ,                                                     \\
         & \hspace{27pt}  \mathfrak{R}\big\{ \widetilde{\mathbf{h}}^H_{n,m,k'} \mathbf{x}_{n,m} \big\} \ge \gamma_{k'}, ~~ \forall k', n, m,                            \\
         & \hspace{27pt}\big| \widetilde{\mathbf{F}}^H \mathbf{x} \big| \leq \sqrt{\frac{P_{\text{T}}}{N_{\text{tot}}}}  \mathbf{1}_{N_{\text{tot}}},
    \end{align}
\end{subequations}
where $\mathbf{m}_t \triangleq \mathbf{g}_t + \widetilde{\mathbf{F}}(\boldsymbol{\lambda} - \rho \mathbf{z})$. It can be seen that problem (\ref{eq:pro_update_x}) is convex and can be solved using various existing methods, such as the interior point method.

\textit{2) Update} $\mathbf{z}$: Given $\mathbf{x}$, $\boldsymbol{\lambda}$ and $\boldsymbol{\mu}$, the optimization problem for updating $\mathbf{z}$ can be formulated as
\begin{equation}\label{eq:pro_update_z}
    \underset{\mathbf{z}}{\min}~~~ \frac{\rho}{2} \big\| \widetilde{\mathbf{F}}^H \mathbf{x} - \mathbf{z} + \frac{\boldsymbol{\lambda}}{\rho} \big\|^2 +\frac{\rho}{2} \big\| |\mathbf{z}| -  \sqrt{\frac{P_{\text{T}}}{N_{\text{tot}}}} \mathbf{1}_{N_{\text{tot}}}  + \frac{\boldsymbol{\mu}}{\rho}  \big\|^2.
\end{equation}
Although (\ref{eq:pro_update_z}) is non-convex due to the absolute value operation, an analytical solution can be obtained in closed form:
\begin{equation}
    \mathbf{z}^{\star} = \frac{1}{2} (|\mathbf{v}| + \mathfrak{R}\{ \mathbf{r} \}) \odot e^{\jmath \angle \mathbf{v}},   \label{eq:up_z}
\end{equation}
where we define
\begin{subequations}
    \begin{align}
        \mathbf{v} & \triangleq \widetilde{\mathbf{F}}^H \mathbf{x} +  \boldsymbol{\lambda} / \rho,                          \\
        \mathbf{r} & \triangleq \sqrt{\frac{P_{\text{T}}}{N_{\text{tot}}}} \mathbf{1}_{N_{\text{tot}}} - \boldsymbol{\mu} / \rho.
    \end{align}
\end{subequations}

\textit{3) Update} $\boldsymbol{\lambda}$ and $\boldsymbol{\mu}$: When both $\mathbf{x}$ and $\mathbf{z}$ are fixed, $\boldsymbol{\lambda}$ and $\boldsymbol{\mu}$ are updated using gradient descent: 
\begin{subequations} \label{eq:up_dual_vari}
    \begin{align}
        \boldsymbol{\lambda}^{\star} & := \boldsymbol{\lambda} + \rho( \widetilde{\mathbf{F}}^H \mathbf{x} - \mathbf{z} ),                            \\
        \boldsymbol{\mu}^{\star}     & := \boldsymbol{\mu} + \rho( |\mathbf{z}| -  \sqrt{\frac{P_{\text{T}}}{N_{\text{tot}}}} \mathbf{1}_{N_{\text{tot}}} ).
    \end{align}
\end{subequations}

\vspace{-5mm}
\subsection{Summary}
Based on the above derivations, the proposed MM-ADMM algorithm using SLP-based ISAC waveform design for range-Doppler sidelobe suppression is summarized in Algorithm \ref{alg:1}, where $\delta_{\text{th}}$ represents the convergence threshold. To expedite the convergence of the proposed algorithm, we adopt the squared iterative method (SQUAREM) \cite{squarem}. The simulation results in Sec. \ref{sec:simulation} demonstrate that the proposed algorithm achieves sufficiently rapid convergence. Next, we discuss how Algorithm \ref{alg:1} should be initialized, and provide an analysis of the required computational complexity.

\textit{1) Initialization :} We propose to initialize $\mathbf{x}$ by maximizing the worst-case communication QoS while satisfying the power budget constraint, which can be formulated as
\begin{subequations}\label{eq:init}
    \begin{align}
         & \underset{\mathbf{x}}{\max}~ \min_{n,m,k'}  \mathfrak{R}\big\{ \widetilde{\mathbf{h}}^H_{n,m,k'} \mathbf{x}_{n,m} \big\}   \\
         & ~\text{s.t.}~~ \big| \widetilde{\mathbf{F}}^H \mathbf{x} \big| \leq \sqrt{\frac{P_{\text{T}}}{N_{\text{tot}}}}  \mathbf{1}_{N_{\text{tot}}}, \label{eq:init_con}
    \end{align}
\end{subequations}
where the relaxed convex power constraint (\ref{eq:init_con}) serves as a substitute for the tight constant-modulus constraint to facilitate finding the initial solution. Problem (\ref{eq:init})  can be efficiently solved as it is convex.

\textit{2) Computational Complexity Analysis:} For this analysis, we assume the commonly used interior point method to solve convex optimization problems, where the computational complexity is related to the dimension of the optimization variable, as well as the number of linear matrix inequality (LMI) and second-order cone (SOC) constraints \cite{Compute}. Thus, the computational complexity of the proposed MM-ADMM algorithm for transmit waveform design mainly arises from the iterative updates of the four variables $\mathbf{x}$, $\mathbf{z}$, $\boldsymbol{\lambda}$, and $\boldsymbol{\mu}$. Given that problem (\ref{eq:pro_update_x}) involves an $N_{\text{tot}}$ dimensional optimization variable with $2 K N_{\text{s}} N_{\text{c}}$ LMI constraints and $N_{\text{tot}}$ SOC constraints, the arithmetic time complexity bound for updating the transmit waveform $\mathbf{x}$ is of order $\mathcal{O} \big\{ \sqrt{(2K + N_{\text{t}})N_{\text{s}}N_{\text{c}}} N_{\text{tot}} (2 N_{\text{tot}}^2 + 2K N_{\text{s}} N_{\text{c}}) \big\}$. The complexity of updating the auxiliary variable $\mathbf{z}$ is of order $\mathcal{O} \big\{ N_{\text{tot}}\}$. Similarly, the complexity for updating the dual variables $\boldsymbol{\lambda}$ and $\boldsymbol{\mu}$ is of order $\mathcal{O} \big\{ N_{\text{tot}} \big\}$. Considering the combined complexities over all iterations, the overall computational complexity of Algorithm \ref{alg:1} can be estimated to be of the order $\mathcal{O} \big\{ \operatorname{ln}(1/\delta_{\text{th}}) \sqrt{(2K + N_{\text{t}})N_{\text{s}}N_{\text{c}}} N_{\text{tot}} (2 N_{\text{tot}}^2 + 2K N_{\text{s}} N_{\text{c}}) \big\}$.

\section{Simulation Results} \label{sec:simulation}
\setlength{\tabcolsep}{0.3 pt}\begin{table}[!t]
    \centering
    \caption{Simulation Parameters}\label{table1}
    \vspace{-2mm}
    {\footnotesize{
            \begin{tabular}{|l|l|l|}
                \hline
                ~\textbf{Parameter }  \hspace{3.5cm}   & ~{\textbf{Symbol}}\hspace{4pt} & ~{\textbf{Value}} \hspace{6pt} \\
                \hline
                ~Carrier frequency~                    & ~$f_{\text{c}}$                & ~24GHz                         \\
                \hline
                ~Number of transmit antennas ~         & ~$N_\text{T}$                  & ~6                             \\
                \hline
                ~Number of receive antennas ~          & ~$N_\text{R}$                  & ~6                             \\
                \hline
                ~Transmit antenna spacing ~            & ~$d_\text{T}$                  & ~$\lambda/2$                   \\
                \hline
                ~Receive antenna spacing ~             & ~$d_\text{R}$                  & ~$\lambda/2$                   \\
                \hline
                ~Number of subcarriers ~               & ~$N_\text{c}$                  & ~32                            \\
                \hline
                ~Number of OFDM symbols ~              & ~$N_\text{s}$                  & ~16                             \\
                \hline
                ~Number of communication users~        & ~$K$                & ~2                             \\
                \hline
                ~Communication QoS requirement~        & ~$\Gamma$                & ~6dB                             \\
                \hline
                ~The communication noise power         & ~$\sigma_{\text{c}}$           & ~-70dBm                         \\
                \hline
                ~The radar noise power                 & ~$\sigma_{\text{r}}$           & ~-70dBm                         \\
                \hline
                ~The power budget for an OFDM frame    & ~$P_\mathrm{T}$                & ~10W                           \\
                \hline
                ~The required minimum illumination power & ~$P_0$                         & ~8W                            \\
                \hline
                ~The convergence threshold             & ~$\delta_{\text{th}}$          & ~$10^{-4}$                     \\
                \hline
            \end{tabular}
        }}
\end{table}

\begin{figure}[!t]
    \centering
    \includegraphics[width = 3.2 in]{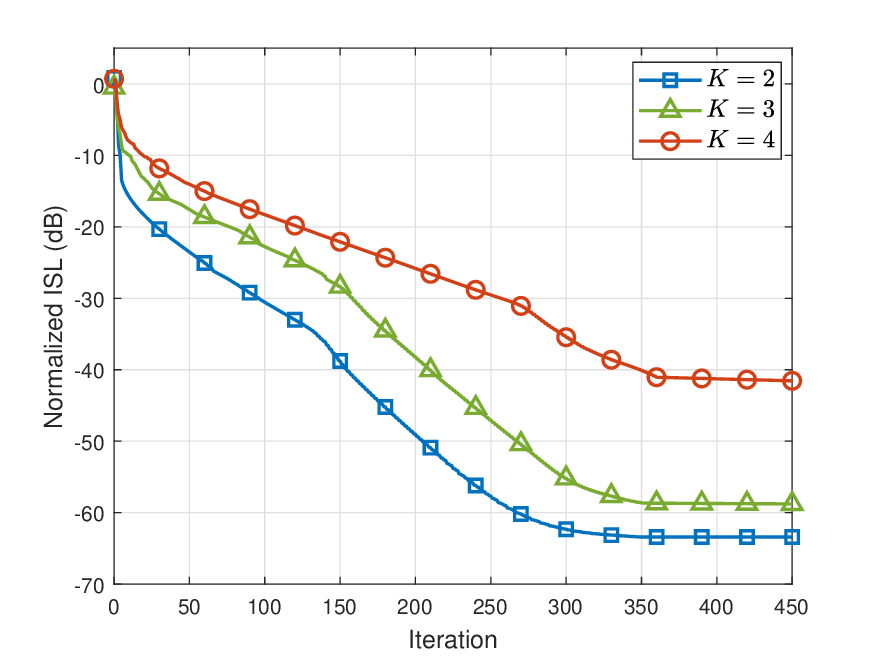}
    \caption{Convergence of the proposed algorithm.}
    \label{fig:convergence} \vspace{-2mm}
\end{figure}

In this section, we showcase simulation results to illustrate the superiority of the proposed MM-ADMM algorithm for the SLP-based ISAC waveform design with range-Doppler sidelobe suppression. Unless otherwise stated, the essential simulation parameters are provided in Table \ref{table1}, where the communication QoS requirement for each user is identical, i.e., $\Gamma = \Gamma_{k}$. The path loss is modeled as $\text{PL}(d) = \zeta_0 (d/d_0)^{-\varepsilon}$, where $\zeta_0 = -30$dB is the path loss at the reference distance $d_0 = 1$m. The distance $d$ between the communication user and the BS is randomly generated between $30$m and $100$m. For the sake of comparison, we also evaluate the performance using the classic dual-functional waveform design method \cite{radar_beam}, which combines the random communication signal and deterministic radar signal using linear block level beamforming. The performance of this benchmark method is referred to as ``\textbf{combined waveform}''.
Moreover, we also include the performance achieved by waveform designs that are exclusively focused on either the communication function or radar function only, and we refer to them as ``\textbf{comm-only waveform}'' and ``\textbf{radar-only waveform}.'' The sidelobe performance of these two methods respectively serve as an upper- and lower-bound for the performance of the proposed algorithm. The comm-only waveform can be easily obtained by solving the max-min fairness problem~(\ref{eq:init}). The radar-only waveform is derived by minimizing the range-Doppler ISL while adhering to the constant-modulus constraint, which can be formulated as
\begin{subequations}	\label{pro_radar}
    \begin{align}
         & \underset{\mathbf{x}}{\min} ~~~  \text{vec}^H(\widetilde{\mathbf{A}}^H \mathbf{x} \mathbf{x}^H \widetilde{\mathbf{A}}) \mathbf{B} \text{vec}(\widetilde{\mathbf{A}}^H \mathbf{x} \mathbf{x}^H \widetilde{\mathbf{A}}) \label{eq:radar_obj} \\
         & ~ \text{s.t.} \hspace{12pt}  \big| \widetilde{\mathbf{F}}^H \mathbf{x} \big| = \sqrt{\frac{P_{\text{T}}}{N_{\text{tot}}}} \mathbf{1}_{N_{\text{tot}}} \label{eq:radar_con}.
    \end{align}
\end{subequations}
We use the MM method with the surrogate upper bound for the range-Doppler ISL in~(\ref{eq:ISL_linear}) to solve~(\ref{pro_radar}). The update of $\mathbf{x}$ can be obtained by solving the following problem
\begin{subequations}	\label{pro_radar_mm}
    \begin{align}
         & \underset{\mathbf{x}}{\min} ~~~  \| \widetilde{\mathbf{F}}^H  (\mathbf{x} + \mathbf{g}_t) \|^2                           \\
         & ~ \text{s.t.} \hspace{12pt}  \big| \widetilde{\mathbf{F}}^H \mathbf{x} \big| = \sqrt{\frac{P_{\text{T}}}{N_{\text{tot}}}} \mathbf{1}_{N_{\text{tot}}},
    \end{align}
\end{subequations}
whose closed-form solution is given by
\begin{equation}
    \mathbf{x} = \sqrt{\frac{P_{\text{T}}}{N_{\text{tot}}}} \widetilde{\mathbf{F}} e^{\jmath \angle ( - \widetilde{\mathbf{F}}^H \mathbf{g}_t )}.
\end{equation}

\begin{figure}[!t]
    \centering
    \subfigbottomskip=-4pt
    \subfigcapskip=-2pt
    \subfigure[The proposed waveform.]{
        \includegraphics[width=0.49\linewidth]{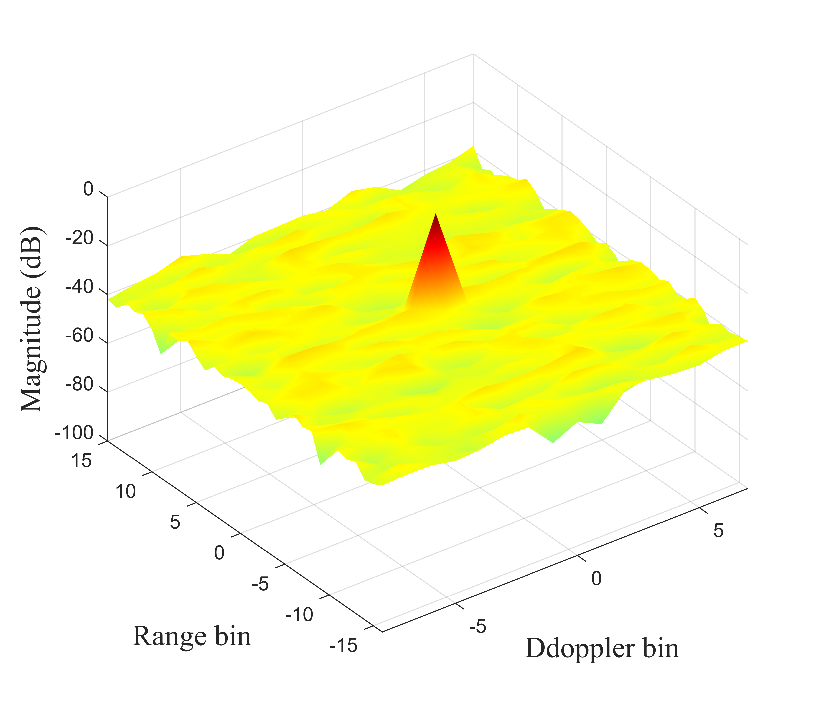} \label{fig:AF_isac}
    }
    \hspace{-8mm}
    \subfigure[The combined waveform.]{
        \includegraphics[width=0.49\linewidth]{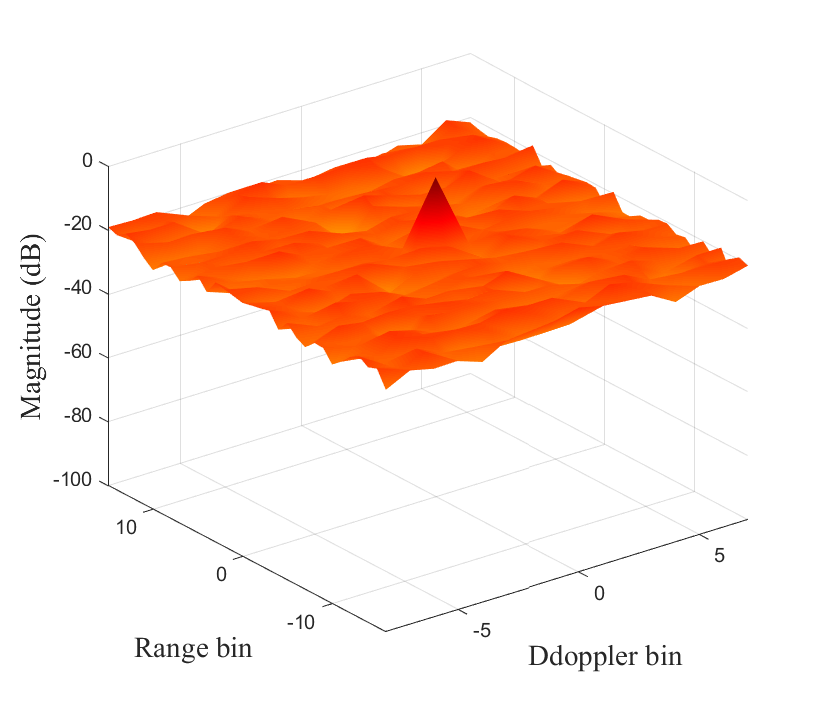} \label{fig:AF_sdr}
    }
    \subfigure[The comm-only waveform.]{
        \includegraphics[width=0.49\linewidth]{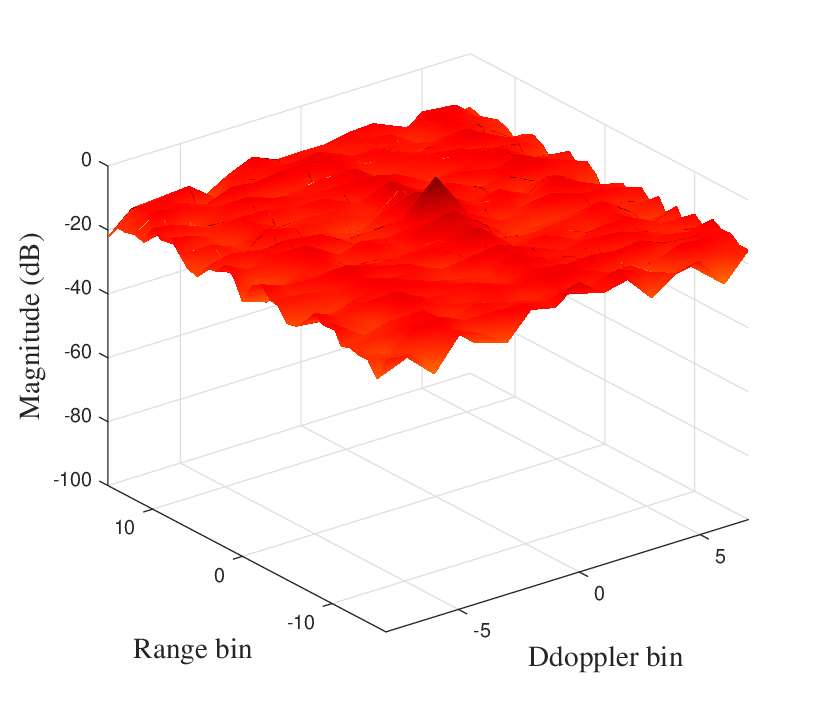} \label{fig:AF_comm}
    }
    \hspace{-8mm}
    \subfigure[The radar-only waveform.]{
        \includegraphics[width=0.49\linewidth]{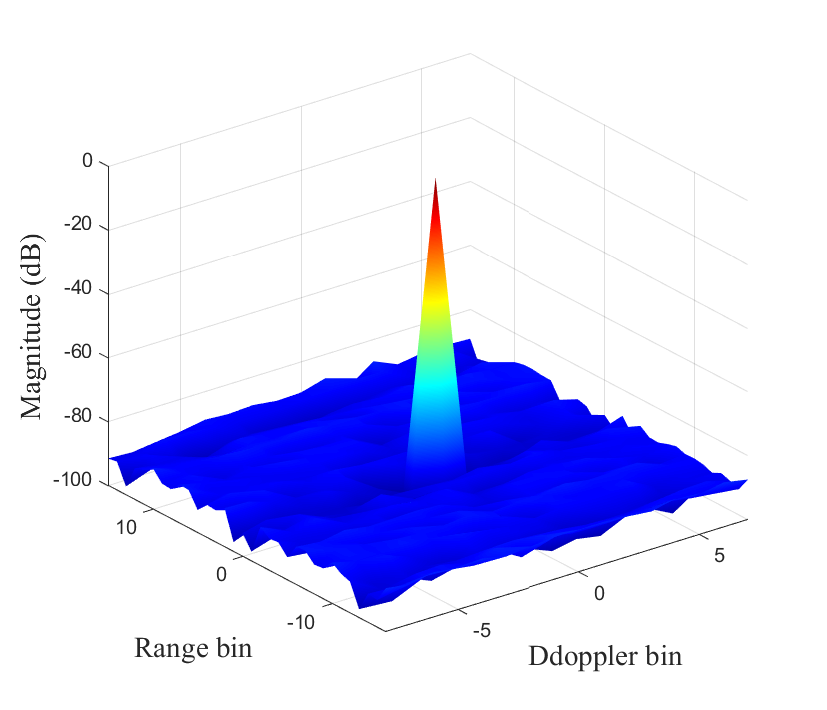} \label{fig:AF_radar}
    }
    \centering
    \vspace{4pt}
    \caption{The ambiguity functions of different waveforms.}\label{fig:ambiguity_function}
\end{figure}

\begin{figure}[!t]
    \centering
    \includegraphics[width = 3.4 in]{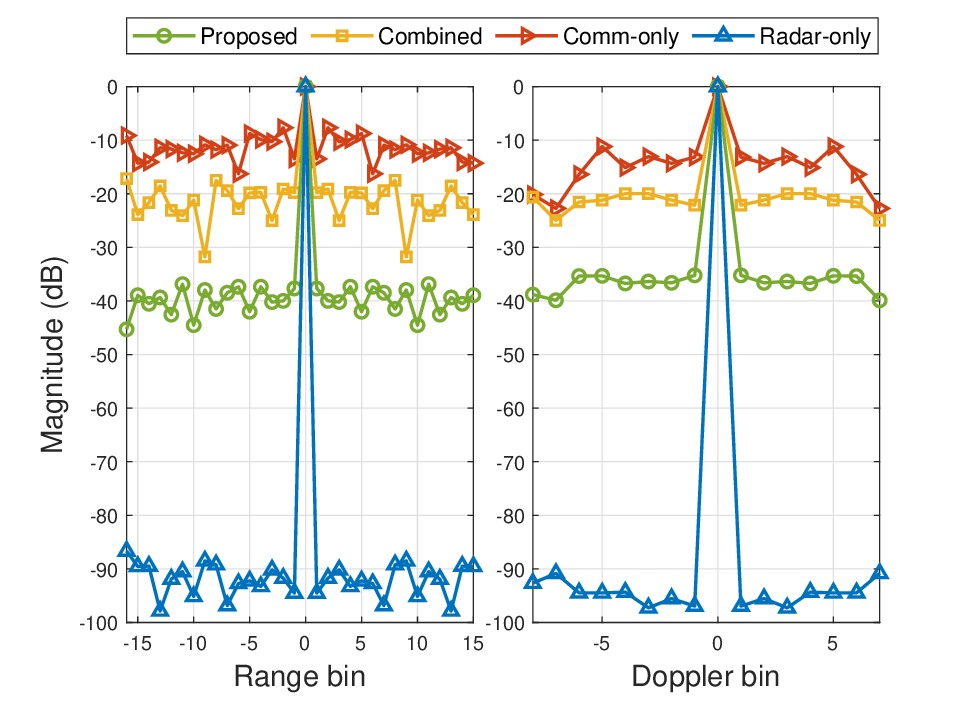}
    \vspace{-2pt}
    \caption{The zero-Doppler and zero-delay slices of the ambiguity functions for different waveform designs.}
    \label{fig:range_Doppler_slice}
\end{figure}

\begin{figure}[!t]
    \centering
    \includegraphics[width = 3.2 in]{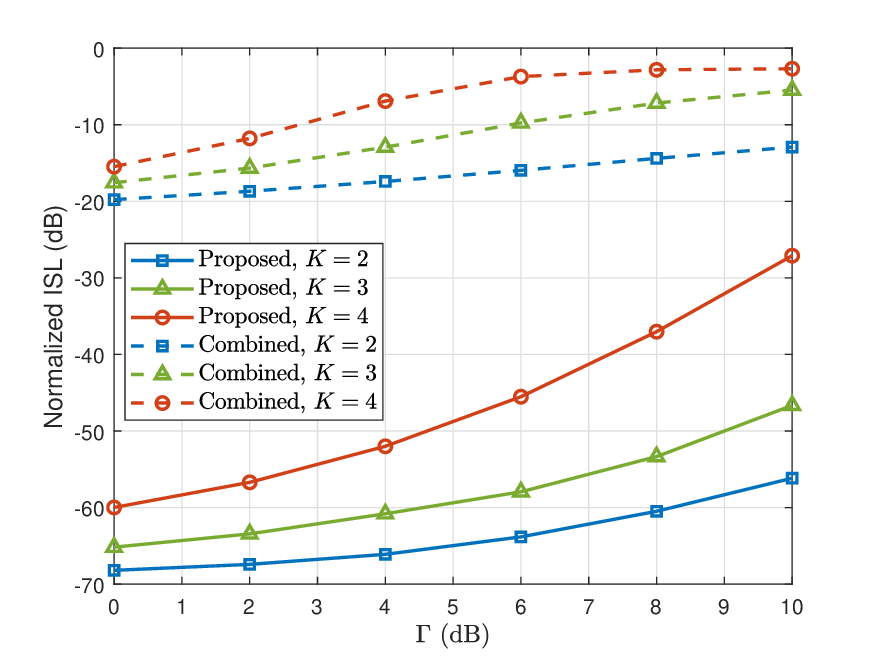}
    \caption{The normalized range-Doppler ISL for different waveform designs versus the communication QoS requirement $\Gamma$, where $N_{\text{c}} = 16$ and $N_{\text{s}} = 4$.}
    \label{fig:ISL_Gamma}
    \vspace{-6pt}
\end{figure}

\begin{figure}[!t]
    \centering
    \subfigbottomskip=-2pt
    \subfigcapskip=-2pt 
    \subfigure[The proposed waveform.]{
        \includegraphics[width=0.49\linewidth]{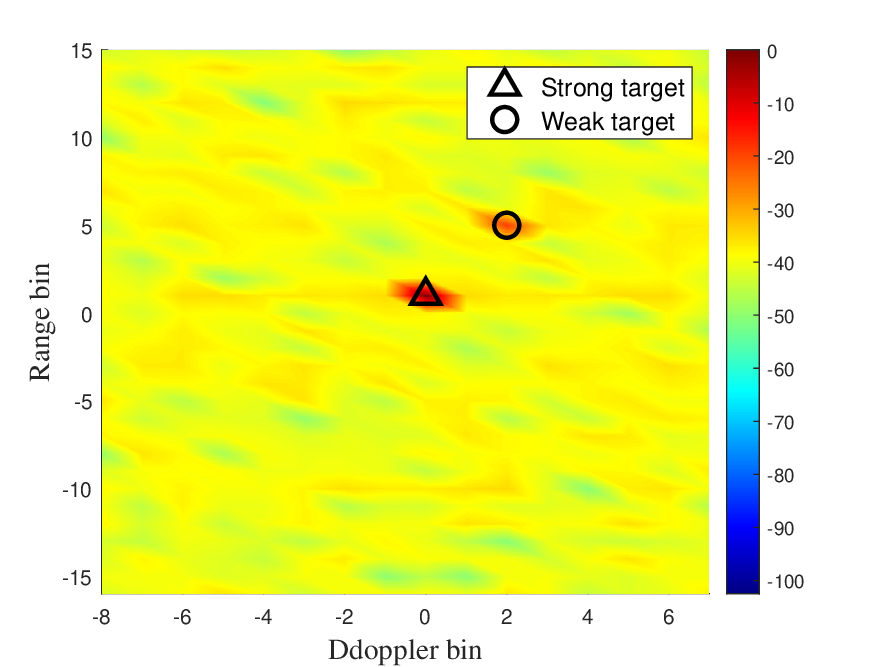}   \label{fig:RDM_isac}
    }
    \hspace{-8mm}
    \subfigure[The combined waveform.]{
        \includegraphics[width=0.49\linewidth]{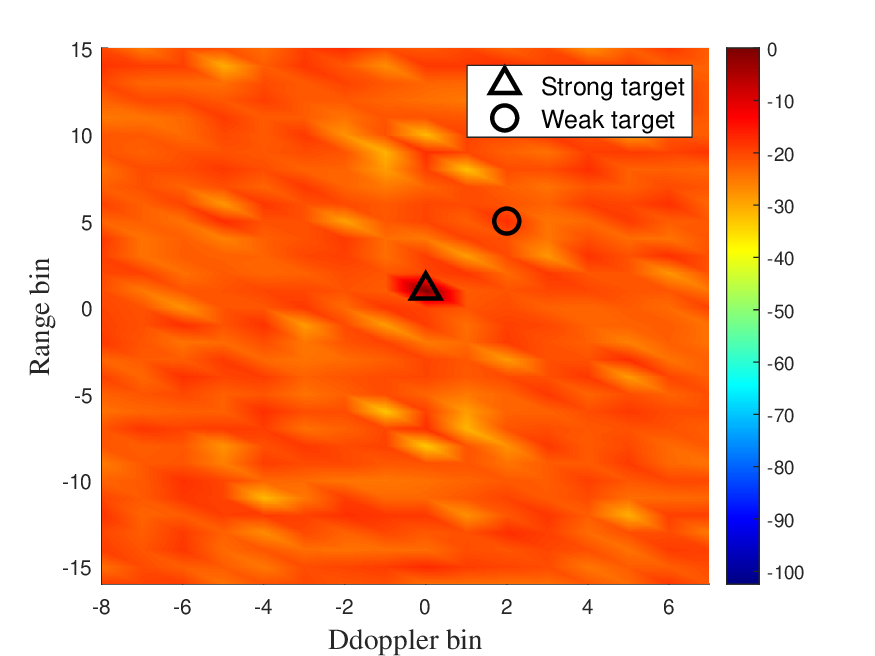}    \label{fig:RDM_sdr}
    }
    \subfigure[The comm-only waveform.]{
        \includegraphics[width=0.49\linewidth]{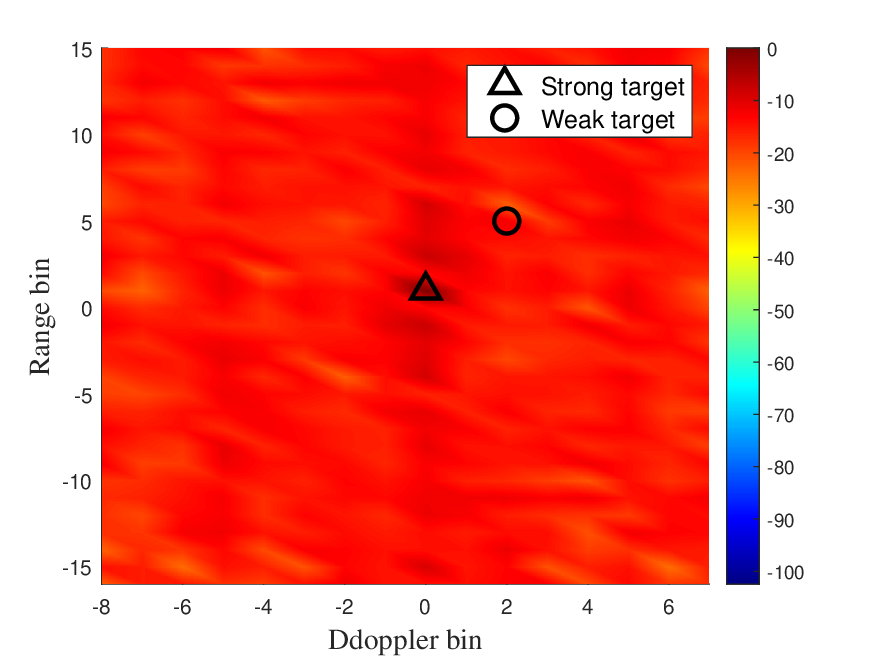} \label{fig:RDM_comm}
    }
    \hspace{-8mm}
    \subfigure[The radar-only waveform.]{
        \includegraphics[width=0.49\linewidth]{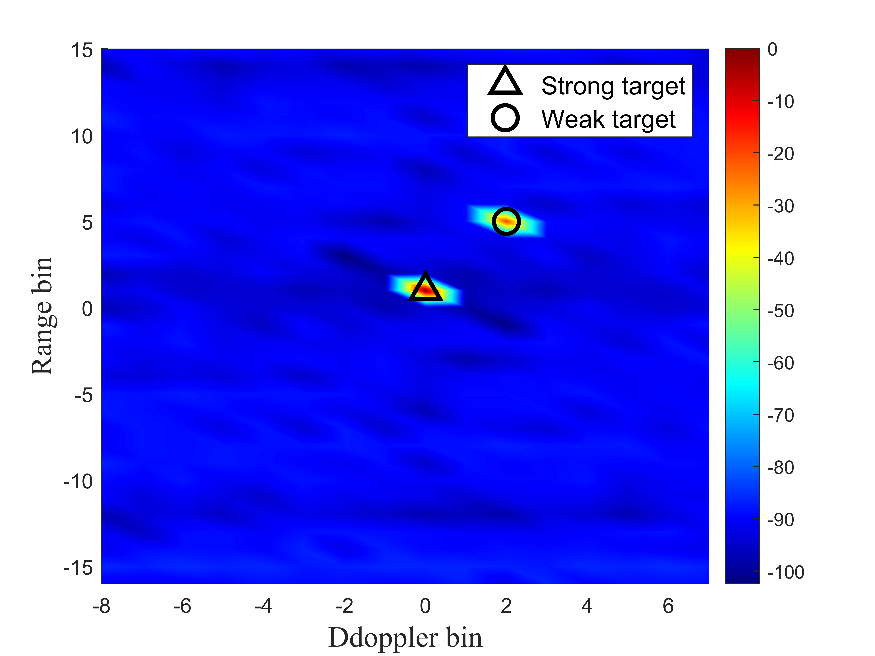} \label{fig:RDM_radar}
    }
    \centering
    \vspace{4pt}
    \caption{Range-Doppler maps for different waveform designs in a scenario with a weak target in close proximity to a strong target.} \label{fig:RDM}
\end{figure}

We first illustrate the average convergence performance of the proposed waveform design algorithm in Fig. \ref{fig:convergence}, which shows a consistent decrease in the ISL with each iteration and convergence within a reasonable number of iterations. 
Next, we evaluate the range-Doppler sidelobe suppression performance using plots of the ambiguity functions of the different waveform designs in Fig.~\ref{fig:ambiguity_function}. We see that the proposed SLP-based waveform exhibits considerably lower range-Doppler sidelobes compared with the approach that uses linear block level beamforming, which validates the advantage of exploiting the extra DoFs offered by the SLP design. We also see that, as predicted, the sidelobe levels of two dual-functional waveforms fall between those of the comm- and radar-only waveforms, which demonstrates the performance trade-off between the two functions.

For a more detailed perspective, the zero-Doppler and zero-delay slices of these ambiguity functions are shown in 
Fig.~\ref{fig:range_Doppler_slice}, clearly illustrating that the proposed SLP-based waveform suppresses range and Doppler sidelobe levels by roughly 15dB more than the combined waveform.
Although the radar-only waveform attains very low range-Doppler sidelobes, it inherently lacks the capability to support the desired communication function. Fig. \ref{fig:ISL_Gamma} evaluates the normalized range-Doppler ISL of the different waveform designs versus the communication QoS requirement. Compared with the combined waveform, the proposed SLP-based waveform significantly reduces the range-Doppler ISL of the ambiguity function for all communication QoS requirements. As expected, we also see that the range-Doppler ISL increases with higher communication QoS requirements and a larger number of communication users, due to the sensing/communications performance trade-off.

Next, we demonstrate the radar sensing performance of the proposed ISAC waveform design in a scenario with two closely-paced targets. Fig.~\ref{fig:RDM} shows the range-Doppler maps of the different algorithms for a strong target with $\sigma_{\text{RCS}}=20$dBsm (e.g., a car) and a weak target with $\sigma_{\text{RCS}}=1$dBsm (e.g., a pedestrian) located at nearby ranges. The comm-only and combined waveforms are unable to identify the weak target since it is submerged in the sidelobes of the strong target. By contrast, for the proposed SLP-based waveform, the mainlobe of the weak target is higher than the sidelobe level of the strong target, which greatly facilitates its detection and estimation even in the presence of the strong target. Through suppression of the range-Doppler sidelobes, the proposed waveform design can effectively enhance the target detection and parameter estimation performance in difficult multi-target environments.

\begin{figure}[!t]
    \centering
    \includegraphics[width = 3.4 in]{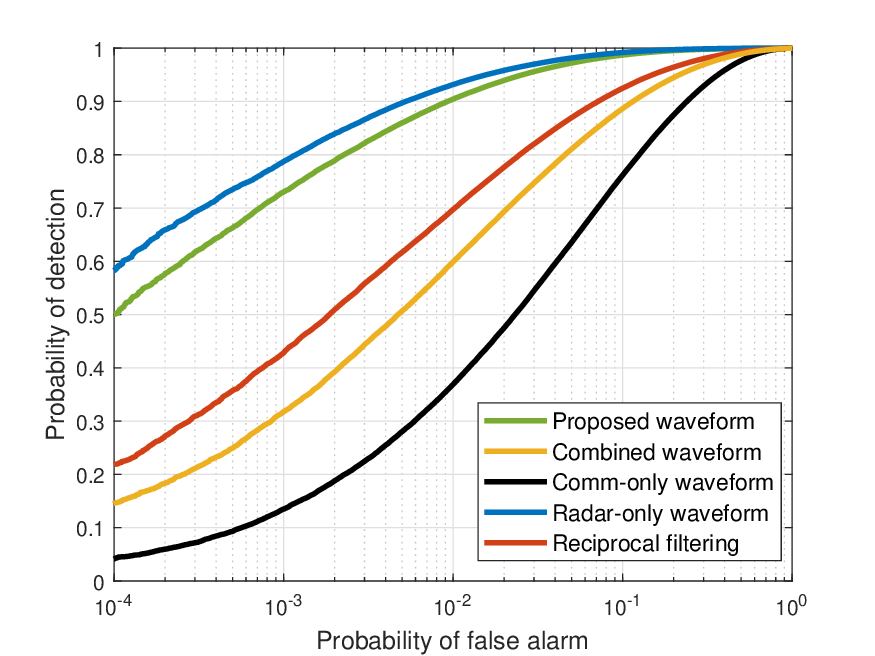}
    \caption{Probability of detection versus probability of false alarm for various algorithms.}
    \label{fig:ROC} \vspace{-1mm}
\end{figure}

\begin{figure}[!t]
    \centering
    \subfigbottomskip=-4pt
    \subfigcapskip=-2pt
    \subfigure[Range estimation RMSE.]{
        \includegraphics[width=3.4 in]{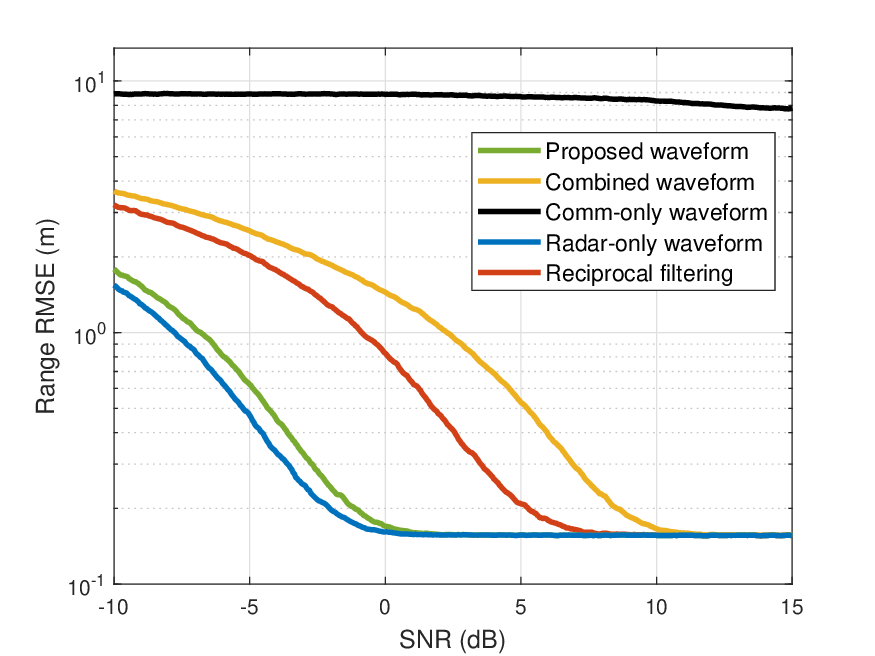} \label{fig:range_RMSE}
    }
    \subfigure[Velocity estimation RMSE.]{
        \includegraphics[width=3.4 in]{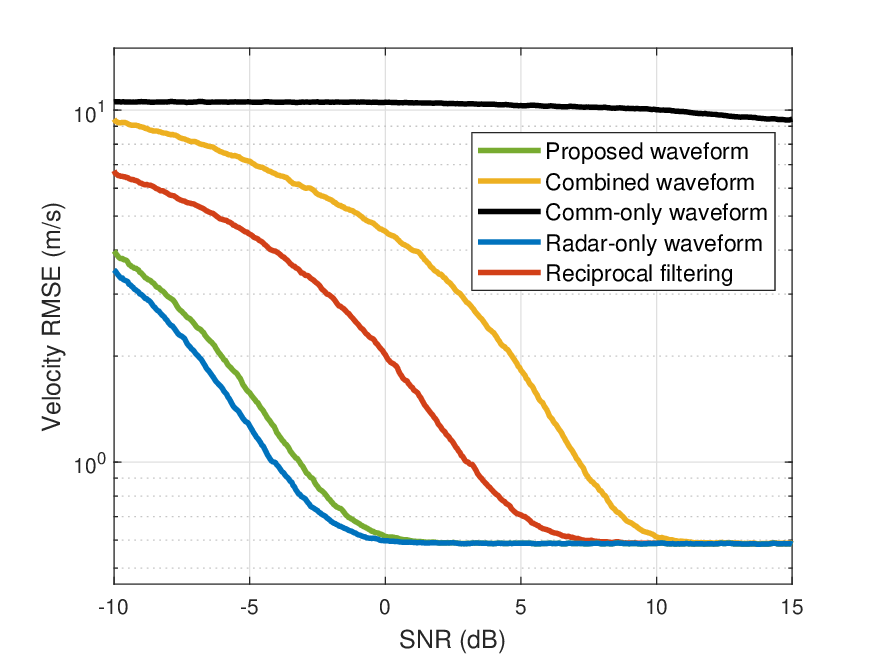} \label{fig:velo_RMSE}
    }
    \centering
    \vspace{4pt}
    \caption{RMSE for range and velocity estimates of the weak target versus the sensing SNR.}\label{fig:RMSE} \vspace{-1mm}
\end{figure}

To evaluate the target detection performance of the proposed waveform design, we depict the receiver operating characteristic (ROC) in Fig.~\ref{fig:ROC}, where the ``\textbf{reciprocal filtering}'' based method \cite{sturm} is also included for comparison. The RCS of the weak target is set as $-3$dBsm and the other parameters remain the same as in Fig. \ref{fig:RDM}. We see that the ROC of the proposed SLP waveform is very close to that for the radar-only case, and both approaches significantly outperform the other benchmarks. Finally, we show the root-mean-square-error (RMSE) of the range and velocity estimates obtained using various waveform designs versus the sensing SNR in Fig.~\ref{fig:RMSE}, where the sensing SNR is defined as the power ratio between the echo of the weak target and the noise. As expected, the comm-only waveform cannot effectively estimate the target parameters due to its high range-Doppler sidelobes. In addition, we see that to achieve identical RMSE performance, the proposed SLP waveform requires 5dB lower SNR gain than reciprocal filtering, and approximately 8dB less than the combined waveform. Again, the proposed SLP design yields almost the same range and velocity RMSE as the radar-only waveform.

\section{Conclusions}\label{sec:conclusion}
\vspace{0.0 cm}
This paper has investigated the advantage of SLP-based ISAC waveform design in MIMO-OFDM ISAC systems, exploiting its significantly increased spatial and temporal DoFs to reduce the range-Doppler sidelobes. We proposed a novel optimization problem to minimize the range-Doppler ISL of the ambiguity function, while satisfying constraints on the target illumination power, the multi-user communication QoS (measured in terms of the gap between the constructive interference region and the symbol decision boundaries), and the constant-modulus of the waveforms. An efficient MM-ADMM-based algorithm was developed to solve the resulting non-convex waveform design problem. Extensive simulation results demonstrated the superiority of the proposed SLP-based ISAC waveform design. The proposed approach achieves performance approaching that of the radar-only design, and much lower range-Doppler sidelobes and better detection and estimation performance than other benchmarks that take the communication requirements into account.

\end{document}